\documentstyle[12pt]{article}
\topmargin - 1.0 true cm
\newcommand{\beq}{\begin{equation}}
\newcommand{\eeq}{\end{equation}}
\def\bea{\begin{eqnarray}}
\def\eea{\end{eqnarray}}
\def\nn{\nonumber}
\def\sss{\scriptscriptstyle}

\def\bd{B_d^0}
\def\bdbar{{\overline{B_d^0}}}
\def\bs{B_s^0}
\def\bsbar{{\overline{B_s^0}}}
\def\barp{{\raise.35ex\hbox
{${\sss (}$}}---{\raise.35ex\hbox{${\sss )}$}}}
\def\bdbarp{\hbox{$B_d$\kern-1.4em\raise1.4ex\hbox{\barp}}}
\def\bsbarp{\hbox{$B_s$\kern-1.4em\raise1.4ex\hbox{\barp}}}
\def\ks{K_{\sss S}}
\def\roughly#1{\mathrel{\raise.3ex\hbox
{$#1$\kern-.75em\lower1ex\hbox{$\sim$}}}}
\def\lsim{\roughly<}

\def\ijmp#1#2#3{{\it Int.\ J.\ Mod.\ Phys.} {\bf A#1} (19#2) #3}

\def\npb#1#2#3{{\it Nucl.\ Phys.} {\bf B#1}, #3 (19#2)}
\def\plb#1#2#3{{\it Phys.\ Lett.} {\bf B#1}, #3 (19#2)}
\def\prd#1#2#3{{\it Phys.\ Rev.} {\bf D#1}, #3 (19#2)}

\def\prl#1#2#3{{\it Phys.\ Rev.\ Lett.} {\bf #1}, #3 (19#2)}
\def\zpc#1#2#3{{\it Zeit.\ Phys.} {\bf C#1}, #3 (19#2)}
\def \stone{{\it B Decays}, edited by S. Stone 
(World Scientific, Singapore, 1994)}
\begin{document}
\baselineskip=6truemm
\begin{flushright}
UdeM-GPP-TH-97-43 \\
YUMS 97-022\\
SNUTP 97-109\\
KEK-TH-532 \\
hep-ph/9708356 \\
\end{flushright}

\begin{center}
\bigskip

{\Large \bf Using $\bs$ Decays to Determine the CP Angles}\\ 
{\Large \bf $\alpha$ and $\gamma$ }\\
\bigskip

C. S. Kim$^{a,}$\footnote{kim@cskim.yonsei.ac.kr,~~
cskim@kekvax.kek.jp},~~ 
D. London $^{b,}$\footnote{london@lps.umontreal.ca}~~ 
and~~ T. Yoshikawa$^{c,}$\footnote{JSPS Research Fellow,~~
yosikawa@theory.kek.jp}
\end{center}


\begin{flushleft}
~~~~~~~~~~~~~$a$: {\it Department of Physics, Yonsei University, 
Seoul 120-749, Korea}\\
~~~~~~~~~~~~~$b$: {\it Laboratoire de physique nucl\'eaire, 
Universit\'e de Montr\'eal,}\\
~~~~~~~~~~~~~~~~~{\it C.P. 6128, succ. centre-ville, Montr\'eal, QC,
Canada}
\\
~~~~~~~~~~~~~$c$: {\it Theory Group, KEK, Tsukuba, Ibaraki 305, Japan}
\end{flushleft}

\begin{center}

\bigskip
(\today)
\bigskip\\

{\bf Abstract}

\end{center}

\begin{quote}
Dighe, Gronau and Rosner have shown that, by assuming $SU(3)$ flavor
symmetry and first-order $SU(3)$ breaking, it is possible to extract the
CP angles $\alpha$ and $\gamma$ from measurements of the decay rates of 
$\bd(t)\to\pi^+\pi^-$, $\bd \to \pi^- K^+$ and $B^+ \to \pi^+ K^0$, along
with their charge-conjugate processes. We extend their analysis to include
the $SU(3)$-related decays $\bs \to \pi^+ K^-$, $\bs(t) \to K^+ K^-$ and
$\bs \to K^0 {\bar K^0}$. There are several advantages to this extension:
discrete ambiguities are removed, fewer assumptions are necessary, and the
method works even if all strong phases vanish. In addition, we show that
$\gamma$ can be obtained cleanly, with no penguin contamination, by using
the two decays $\bs(t) \to K^+ K^-$ and $\bs \to K^0 {\bar K^0}$.
\end{quote}

\newpage


\section{Introduction}

In the coming years, the CP angles $\alpha$, $\beta$ and $\gamma$, which
characterize the unitarity triangle, will be measured at $B$ factories.
Through such measurements it will be possible to test the Standard Model
(SM) explanation of CP violation, namely that CP violation is due to a
complex phase in the Cabibbo-Kobayashi-Maskawa (CKM) matrix. Because of the
importance of such tests, many ingenious ways of getting at the CP angles
have been devised \cite{CPreview}. 

All methods involve CP-violating rate asymmetries between the decays $B \to
f$ and ${\bar B} \to {\bar f}$. The conventional ways of measuring $\alpha$
and $\beta$ use the CP asymmetries in $\bd(t)\to\pi^+\pi^-$ and
$\bd(t)\to\Psi\ks$, respectively, while the CP angle $\gamma$ can be
obtained through the CP asymmetry in the charged $B$ decay $B^\pm \to D\
K^\pm$ \cite{growyler}. Alternatively, $\gamma$ can be measured via
$\bs(t)\to D_s^\pm K^\mp$ \cite{ADK}. 

There are many other decay modes which can give information about the CP
angles. However, one of the problems which must be addressed is the
question of penguin contamination. For example, the penguin contribution
to $\bd \to \pi^+\pi^-$ is likely to be sizeable, so that $\alpha$ cannot
be cleanly extracted from the measurement of the CP asymmetry in
$\bd(t)\to\pi^+\pi^-$. In this case one can remove the effects of the
penguin amplitude by performing an isospin analysis \cite{isospin}, but
this will probably not be easy since it requires the measurement of the
branching ratio of $\bd \to \pi^0\pi^0$, which is expected to be of
order $10^{-6}$ or less.

Recently, Dighe, Gronau and Rosner (DGR) proposed an elegant new method
\cite{GR,DGR,DR} of dealing with this penguin contamination. This method
requires the measurement of the rates for $\bd \to \pi^- K^+$ and $B^+ \to
\pi^+ K^0$ ($\ks \to \pi^+ \pi^-$), as well as their charge-conjugate
processes, along with the CP asymmetry in $\bd(t)\to\pi^+\pi^-$. Using
$SU(3)$ flavor symmetry \cite{SU3} and first-order $SU(3)$ breaking
\cite{SU3break}, these measurements can be used to disentangle the effects
of the penguin contribution, and thus obtain $\alpha$ cleanly. In addition,
the weak phase $\gamma$ can also be extracted from this set of measurements.

There are a number of advantages to this method. First, it does not suffer
from the problems with electroweak penguins \cite{EWPs} and $SU(3)$ breaking
that plague other methods. Second, it uses only decays of $\bd$ and $B^+$
mesons, which are accessible at asymmetric $e^+e^-$ colliders running on
the $\Upsilon(4S)$ ($B$ factories). Finally, the decays involve only
charged $\pi$'s or $K$'s, which makes the measurements considerably easier.

However, there are also some problems with this method. First, there is a
large number of discrete ambiguities in the extraction
of $\alpha$, $\gamma$, and the strong phase difference between the tree and
penguin diagrams, $\delta$. Many of these ambiguities can be rejected due
to other information that we have about the CKM matrix, but some still
remain. This creates some difficulties in identifying the presence of new
physics. Second, there are some theoretical assumptions which, while
reasonable, may turn out not to be true. In particular, DGR assume that
the strong phase of the penguin diagram in $\Delta S=0$ transitions is
equal to that of the penguin diagram in $\Delta S=1$ transitions. They also
assume that the $b\to d$ penguin is dominated by an internal $t$ quark. If
any of these assumptions is relaxed, then there is not enough information
from the measurements to determine $\alpha$ and $\gamma$. Finally, the
method also breaks down if $\delta$ vanishes. In this case it is necessary
to make additional assumptions in order to extract information about the
CP angles.

In this paper, we discuss an extension of the DGR method which eliminates
many of these problems, or at least improves upon them. In addition to the
$\bd$ and $B^+$ decays used by DGR, this extension uses their
$SU(3)$-counterpart $\bs$ decays: $\bs \to \pi^+ K^-$, $\bs(t) \to K^+ K^-$,
and $\bs \to K^0 {\bar K^0}$. If we make the same assumptions as DGR, we
are able to extract $\alpha$ and $\gamma$ with a 2-fold ambiguity,
corresponding to the unitarity triangle pointing up or down. If we relax
the DGR assumptions, we are still able to obtain the CP angles up to
possible discrete ambiguities. Even if $\delta=0$, we are still able to
extract these angles, up to some discrete ambiguities. Thus, this
extension allows us to extract the CP angles $\alpha$ and $\gamma$, up to
possible discrete ambiguities, with a minimum of theoretical assumptions.

If one relaxes all theoretical assumptions, then we find that it is not
possible to extract $\alpha$ from measurements of the six decays. However,
perhaps surprisingly, it is still possible to obtain $\gamma$. We show
that measurements of the decays $\bs(t) \to K^+ K^-$ and $\bs \to K^0 {\bar
K^0}$ alone allow the measurement of $\gamma$ with no hadronic uncertainty. 
This is a new way of obtaining this angle.

The paper is organized as follows. In Section 2 we review the DGR method,
followed in Section 3 by a description of our extension of this method. We
then examine the effects of relaxing the DGR assumptions. The
case $\delta\ne\delta'$ is considered in Section 4, followed by the
inclusion of internal $u$ and $c$ quarks in the $b\to d$ penguins in Section
5. Section 5 also includes the description of a new method for
obtaining $\gamma$. In Sections 3-5 it is assumed that the strong phases
are independent of the spectator quark. Section 6 discusses the case where
this assumption is relaxed. In Section 7 we consider the case of vanishing
strong phases. We conclude in Section 8.


\section{The DGR Method}

In this section we review the method of Dighe, Gronau and Rosner
\cite{GR,DGR,DR}. Using $SU(3)$ symmetry \cite{SU3}, all $B \to PP$ decays,
where $B$ represents $\bd$, $B^+$ or $\bs$, and $P$ is a pseudoscalar meson, can
be written in terms of five $SU(3)$ amplitudes. These five $SU(3)$
amplitudes can in turn be expressed in terms of six diagrams. Of these six
diagrams, three of them --- the exchange, annihilation, and penguin
annihilation diagrams --- can be neglected, since they are expected to be
suppressed by $f_{\sss B}/m_{\sss B} \sim$ 3-4\%. The remaining amplitudes
are the tree $t$ ($t'$), color-suppressed $c$ ($c'$), and penguin $p$ ($p'$)
terms, where the unprimed and primed quantities denote $\Delta S=0$
and $\Delta S=1$ processes, respectively. These amplitudes include both the
leading-order and electroweak penguin \cite{EWPs} contributions:
\bea
t & \equiv & T + (c_u - c_d) P_{\sss EW}^{\sss C} \nn \\
c & \equiv & C + (c_u - c_d) P_{\sss EW} \\
p & \equiv & P + c_d P_{\sss EW}^{\sss C}~, \nn 
\eea
where $P_{\sss EW}$ and $P_{\sss EW}^{\sss C}$ are the color-favored and
color-suppressed electroweak penguin (EWP) diagrams, respectively,
and $c_u$ and $c_d$ are the couplings of the $Z$ to $u$ quarks and $d$ quarks,
respectively. In fact, although the EWP contributions have been included
above, in most processes they are at the level of exchange-type diagrams,
and so are negligible. Only the color-favored EWP is non-negligible, and
then only in $\Delta S=1$ transitions. 

The DGR method involves the decays $\bd \to\pi^+\pi^-$, $\bd \to \pi^-
K^+$, and $B^+ \to \pi^+ K^0$. The amplitudes for these decays can be
written
\bea
\label{PPamps}
A_{\pi\pi} &\equiv & A\left(B^0\rightarrow \pi^+\pi^-\right) = 
 - \left( T + P \right), \nn \\
A_{\sss \pi K} &\equiv & A\left(B^0\rightarrow \pi^- K^+\right) = 
 - \left( T^\prime + P^\prime \right), \\
A_{\sss \pi K}^+ &\equiv & A\left(B^+\rightarrow\pi^+K^0\right) = 
 P^\prime ~.
\nn
\eea
Note that, since only tree and penguin terms are involved, EWP
contributions are negligible. (In fact, DGR include the EWP contributions,
but end up effectively setting them to zero by making the approximation
$-{1\over 3} P_{\sss EW}^{\sss \prime C} \approx {2\over 3} P_{\sss
EW}^{\sss \prime C}$.)

The weak phase of $T$ is Arg$\left(V_{ud} V_{ub}^* \right) = \gamma$, and
similarly for $T'$: Arg$\left(V_{us} V_{ub}^* \right) = \gamma$. The $b\to
s$ penguin $P'$ is dominated by the internal $t$-quark, so its weak phase is
Arg$\left(V_{ts} V_{tb}^* \right) = \pi$. As for the $b\to d$ penguin $P$, 
if it also is dominated by the $t$-quark, its weak phase is Arg$\left(V_{td}
V_{tb}^* \right) = - \beta$. This is the assumption made by DGR, but we
will relax it in later sections.

If $SU(3)$ were unbroken, the amplitudes $T$ and $T'$ would be related simply
by the ratio of their CKM matrix elements: $|T'/T| = |V_{us}/V_{ud}|$.
However, if one includes first-order $SU(3)$ breaking \cite{SU3break}, there
is an additional factor involving the ratio of $K$ and $\pi$ decay constants
if factorization is assumed:
\beq
\left|\frac{T^\prime}{T} \right| = \frac{\left|V_{us}\right| f_{\sss K}} 
 { \left| V_{ud} \right| f_\pi } \equiv r_u ~.
\eeq
On the other hand, since factorization is unlikely to hold for penguin
amplitudes, $P$ and $P'$ are not related in a simple way. However, DGR do
assume that the strong phase of the penguin diagram, $\delta_{\sss P}$, is
unaffected by $SU(3)$ breaking. This assumption will also be relaxed in
later sections.

With these assumptions, the amplitudes in Eqs.~(\ref{PPamps}) can be
written
\bea
A_{\pi\pi} & = & {\cal T} e^{i \delta_T} e^{i\gamma } +
 {\cal P} e^{i \delta_P} e^{- i\beta } ~, \nn \\
A_{\sss \pi K} & = & r_u {\cal T} e^{i \delta_T} e^{i\gamma } -
 {\cal P}^\prime e^{i \delta_P} ~, \\
A_{\sss \pi K}^+ & = & {\cal P}^\prime e^{i \delta_P} ~, \nn
\eea
where ${\cal T} \equiv \left| T \right|$, ${\cal P} \equiv \left| P
\right|$, and ${\cal P}^\prime \equiv \left| P^\prime \right|$.

There are thus six unknown quantities in the above 3 amplitudes: $\alpha
\equiv \pi - \beta - \gamma$, $\gamma$, ${\cal T}$, ${\cal P}$, ${\cal
P}^\prime$, and $\delta \equiv \delta_{\sss T} - \delta_{\sss P}$. These
quantities can be extracted as follows. The time-dependent, tagged $\bd$
and $\bdbar$ decay rates to $\pi^+\pi^-$ are given by
\bea
\Gamma\left( \bd(t) \rightarrow \pi^+ \pi^- \right) &=& 
 e^{-\Gamma t} \left[ 
 \left| A_{\pi \pi } \right|^2 \cos^2 
 \left( \frac{\Delta m}{2} t 
 \right) +
 \left| \bar{A}_{\pi \pi } \right|^2 \sin^2 
 \left( \frac{\Delta m}{2} t 
 \right) 
 \right. \nn \\ 
 \ &\ & \hspace{4.5cm} + \left. 
 {\rm Im}\left(e^{2i\beta }A_{\pi \pi }
 \bar{A}_{\pi \pi }^* \right)
 \sin(\Delta mt)
 \right], \nn \\ 
\Gamma\left( \bdbar(t) \rightarrow \pi^+ \pi^- \right) &=& 
 e^{-\Gamma t} \left[ 
 \left| A_{\pi \pi } \right|^2 \sin^2 
 \left( \frac{\Delta m}{2} t 
 \right) +
 \left| \bar{A}_{\pi \pi } \right|^2 \cos^2 
 \left( \frac{\Delta m}{2} t 
 \right) 
 \right. \nn \\
 \ &\ & \hspace{4.5cm} - \left. 
 {\rm Im}\left(e^{2i\beta }A_{\pi \pi }
 \bar{A}_{\pi \pi }^* \right)
 \sin(\Delta mt)
 \right]. 
\eea
{}From these measurements one can determine the three quantities $\left|
A_{\pi \pi } \right|^2$, $\left| \bar{A}_{\pi \pi } \right|^2$, and ${\rm
Im}\left(e^{2i\beta} A_{\pi\pi} \bar{A}_{\pi \pi }^*\right)$. The rates
for the self-tagging decays $\bd \to \pi^- K^+$ and $\bdbar \to \pi^+ K^-$
are
\bea
|A_{\sss \pi K}|^2 & = & r_u^2 {\cal T}^2 + 
{{\cal P}^\prime}^2 - 2 r_u {\cal T}
{\cal P}^\prime \cos (\delta + \gamma), \nn \\
|{\bar A}_{\sss \pi K}|^2 & = & r_u^2 {\cal T}^2 + 
{{\cal P}^\prime}^2 - 2 r_u
{\cal T} {\cal P}^\prime \cos (\delta - \gamma).
\eea
Finally, the rates for $B^+ \to \pi^+ K^0$ and its CP-conjugate decay give
\beq
|A_{\sss \pi K}^+|^2=|{\bar A}_{\sss \pi K}^-|^2={{\cal P}^\prime}^2 ~.
\eeq

Thus, from the above measurements, one can obtain the following six
quantities:
\bea
A &\equiv & \frac{1}{2}
 \left( 
 \left| A_{\pi \pi }\right|^2 
 + \left| \bar{A}_{\pi \pi }\right|^2 
 \right) = 
 {\cal T}^2 + {\cal P}^2 
 - 2 {\cal T}{\cal P} \cos\delta \cos\alpha, 
\label{A1} \\
B &\equiv & \frac{1}{2}
 \left( 
 \left| A_{\pi \pi }\right|^2 
 - \left| \bar{A}_{\pi \pi }\right|^2 
 \right) = 
 - 2 {\cal T}{\cal P} \sin\delta \sin\alpha, 
\label{B1} \\
C &\equiv & {\rm Im}\left( e^{2 i \beta} A_{\pi \pi } 
 \bar{A}_{\pi \pi }^* 
 \right)
 = - {\cal T}^2 \sin2\alpha 
 + 2 {\cal T P} \cos\delta \sin\alpha, 
\label{C1} \\
D &\equiv & \frac{1}{2}
 \left( 
 \left| A_{\sss \pi K}\right|^2 
 + \left| \bar{A}_{\sss \pi K}\right|^2 
 \right) = 
 r_u^2 {\cal T}^2 + {{\cal P}^\prime}^2 
 - 2 r_u {\cal T}{\cal P}^\prime
 \cos\delta \cos\gamma, 
\label{D1} \\
E &\equiv & \frac{1}{2}
 \left( 
 \left| A_{\sss \pi K}\right|^2 
 - \left| \bar{A}_{\sss \pi K}\right|^2 
 \right) = 
 2 r_u {\cal T}{\cal P}^\prime
 \sin\delta \sin\gamma, 
\label{E1} \\ 
F &\equiv & \left| A_{\sss \pi K}^+\right|^2 = {{\cal P}^\prime}^2~.
\label{F1}
\eea
These give 6 equations in 6 unknowns, so that one can solve for 
$\alpha$, $\gamma$, ${\cal T}$, ${\cal P}$, ${\cal P}^\prime$, and $\delta$.
However, because the equations are nonlinear, there are discrete
ambiguities in extracting these quantities. In fact, a detailed study
\cite{DR} shows that, depending on the actual values of the phases, there
can be up to 8 solutions. Many of these can be eliminated due to other
information on the CKM phases, but still some ambiguity often remains.

{}From this brief summary, one can see some of the problems of the method. 
If $\delta=0$, the quantities $B$ and $E$ vanish, so that one is left with 4
equations in 5 unknowns. In this case one must use additional assumptions
to extract information about the CP phases. Furthermore, even if $\delta
\ne 0$, if one relaxes any of the assumptions described above, the method
breaks down. For example, if one allows the strong phase of the $P'$ diagram
to be different from that of the $P$ diagram, as might be the case in the
presence of $SU(3)$ breaking, then one has 6 equations in 7 unknowns. And
if one relaxes the assumption that the $b\to d$ penguin is dominated by
the $t$-quark, then once again additional parameters are introduced, and
the method breaks down.

All of these potential problems can be dealt with by considering
additional $B_s^0$ decays. We discuss this possibility in the following
sections.


\section{Extending the DGR Method with $\bs$ Decays}

The problems with the DGR method can be resolved by adding amplitudes
which depend on the same 6 quantities, thus overconstraining the system. In
this case, if one adds a parameter or two, perhaps by relaxing certain
assumptions, the method will be less likely to break down. 

Within $SU(3)$ symmetry, the obvious decays to consider are the $SU(3)$
counterparts to the DGR decays, namely $\bs \to \pi^+ K^-$, $\bs(t) \to K^+
K^-$, and $\bs \to K^0 {\bar K^0}$. The amplitudes for these decays are
completely analogous to those in Eqs.~(\ref{PPamps}):
\bea
\label{BsPPamps}
B_{\sss \pi K} &\equiv & A\left(\bs\rightarrow \pi^+ K^-\right) = 
 - \left( \tilde{T} + \tilde{P} \right), \nn \\
B_{\sss KK} &\equiv & A\left(\bs\rightarrow K^+ K^-\right) = 
 - \left( {\tilde T}^\prime + \tilde{P}^\prime \right), \\ 
B_{\sss KK}^s &\equiv & A\left(\bs\rightarrow K^0{\overline{K^0}} 
\right) = \tilde{P}^\prime ~.
\nn
\eea
Here we have denoted the tree and penguin diagrams involving a spectator
$s$ quark by $\tilde{T}$ and $\tilde{P}$, respectively. As before, the
unprimed and primed quantities denote $\Delta S=0$ and $\Delta S=1$
processes, respectively. 

The weak phase of $\tilde{T}$ and $\tilde{T}^\prime$ is $\gamma$, and that
of $\tilde{P}^\prime$ is $\pi$. As for $\tilde{P}$, as a first step we make
the same assumptions as DGR, namely that it is dominated by the $t$-quark,
so that its weak phase is $-\beta$. Turning to $SU(3)$ breaking, we assume
factorization for the tree amplitudes, so that
\beq
\left| \frac{\tilde{T}^\prime}{\tilde{T}} \right| = r_u ~.
\eeq
The magnitudes of the $\tilde{P}$ and $\tilde{P}^\prime$ amplitudes are
unrelated to one another. However, again as a first step, like DGR we
assume that they have the same strong phase, $\delta_{\sss\tilde{P}}$. In
subsequent sections, we will examine the consequences of relaxing these
assumptions.

The one new assumption that we make is that the relative strong phase
between the tree and penguin amplitudes is independent of the flavor of the
spectator quark. Thus we have $\delta_s = \delta$, where $\delta_s \equiv
\delta_{\sss\tilde{T}} - \delta_{\sss\tilde{P}}$ and $\delta \equiv
\delta_{\sss T} - \delta_{\sss P}$. (The most likely way for this to occur
is if $\delta_{\sss T} = \delta_{\sss\tilde{T}}$ and $\delta_{\sss P} =
\delta_{\sss\tilde{P}}$.) This assumption, which is motivated by the
spectator model, will also be reexamined in later sections.

Under these assumptions, the amplitudes in Eqs.~(\ref{BsPPamps}) can be
written
\bea
B_{\sss \pi K} & = & \tilde{{\cal T}} e^{i \delta_T} e^{i\gamma } +
 \tilde{{\cal P}} e^{i \delta_P} e^{- i\beta } ~, \nn \\
B_{\sss KK} & = & r_u \tilde{{\cal T}} e^{i \delta_T} e^{i\gamma } -
 \tilde{{\cal P}}^\prime e^{i \delta_P} ~, \\
B_{\sss KK}^s & = & \tilde{{\cal P}}^\prime e^{i \delta_P} ~, \nn
\eea
where $\tilde{\cal T} \equiv \left| \tilde{T} \right|$, $\tilde{\cal P}
\equiv \left| \tilde{P} \right|$, and $\tilde{\cal P}^\prime \equiv \left|
\tilde{P}^\prime \right|$.

The important point here is that three new parameters have been introduced
in the above amplitudes: $\tilde{{\cal T}}$, $\tilde{{\cal P}}$, and
$\tilde{{\cal P}}^\prime$. However, as in the DGR method, 6 quantities can
be extracted from measurements of the rates for these decays. Here, the
self-tagging decays are $\bs \to \pi^+ K^-$ and $\bsbar \to \pi^- K^+$,
whose rates are
\bea
|B_{\sss \pi K}|^2 & = & {\tilde{\cal T}}^2 + 
{\tilde{\cal P}}^2 - 2 \tilde{\cal T} \tilde{\cal P} 
\cos (\delta - \alpha), \nn \\
|{\bar B}_{\sss \pi K}|^2 & = & 
{\tilde{\cal T}}^2 + {\tilde{\cal P}}^2 - 2
\tilde{\cal T} \tilde{\cal P} \cos (\delta + \alpha).
\eea
The time-dependent, tagged $\bs$ and $\bsbar$ decay rates to $K^+K^-$ are
given by 
\bea
\Gamma\left[ \bs(t) \rightarrow K^+ K^- \right] &=& 
 e^{-\Gamma t} \left[ 
 \left| B_{\sss KK} \right|^2 \cos^2 
 \left( \frac{\Delta m_s}{2} t 
 \right) +
 \left| \bar{B}_{\sss KK} \right|^2 \sin^2 
 \left( \frac{\Delta m_s}{2} t 
 \right) 
 \right. \nn \\
 \ &\ & \hspace{4.5cm} + \left. 
 {\rm Im}\left( B_{\sss KK}
 \bar{B}_{\sss KK}^* \right)
 \sin(\Delta m_s t)
 \right], \nn \\ 
\Gamma\left[ \bsbar(t) \rightarrow K^+ K^- \right] &=& 
 e^{-\Gamma t} \left[ 
 \left| B_{\sss KK} \right|^2 \sin^2 
 \left( \frac{\Delta m_s}{2} t 
 \right) +
 \left| \bar{B}_{\sss KK} \right|^2 \cos^2 
 \left( \frac{\Delta m_s}{2} t 
 \right) 
 \right. \nn \\ 
 \ &\ & \hspace{4.5cm} - \left. 
 {\rm Im}\left( B_{\sss KK}
 \bar{B}_{\sss KK}^* \right)
 \sin(\Delta m_s t)
 \right], 
\eea
from which the quantities $\left| B_{\sss KK} \right|$, $\left|
\bar{B}_{\sss KK} \right|$, and ${\rm Im}\left( B_{\sss KK}\bar{B}_{\sss
KK}^* \right)$ can be extracted. Finally, we turn to $\bs(t) \to K^0
{\overline{K^0}}$. In principle there can be indirect CP violation in
these decays. However, within the SM, this CP violation is zero to a good
approximation, since both $\bs$-$\bsbar$ mixing and the $b\to s$ penguin
diagram, which dominates this decay, are real. Thus, measurements of the
rates for these decays yield
\beq
|B_{\sss KK}^s|^2 = |{\bar B}_{\sss KK}^s|^2 = 
{{\tilde{\cal P}}^\prime}{}^2 ~.
\eeq
Obviously, any violation of this equality will be clear evidence for new
physics.

Therefore the above measurements yield 6 new quantities:
\bea
\tilde{A} &\equiv & \frac{1}{2}
 \left( 
 \left| B_{\sss \pi K}\right|^2 
 + \left| \bar{B}_{\sss \pi K}\right|^2 
 \right) = 
 \tilde{\cal T}^2 + \tilde{\cal P}^2 
 - 2 \tilde{\cal T}
 \tilde{\cal P} \cos\delta \cos\alpha, 
\label{A2} \\
\tilde{B} &\equiv & \frac{1}{2}
 \left( 
 \left| B_{\sss \pi K}\right|^2 
 - \left| \bar{B}_{\sss \pi K}\right|^2 
 \right) = 
 - 2 \tilde{\cal T}\tilde{\cal P} 
 \sin\delta \sin\alpha, 
\label{B2} \\
\tilde{C} &\equiv & {\rm Im}\left( B_{\sss KK} 
 \bar{B}_{\sss KK}^* 
 \right)
 = r_u^2 \tilde{\cal T}^2 \sin2\gamma 
 - 2 r_u \tilde{\cal T }\tilde{\cal P}^\prime 
 \cos\delta \sin\gamma, 
\label{C2} \\ 
\tilde{D} &\equiv & \frac{1}{2}
 \left( 
 \left| B_{\sss KK}\right|^2 
 + \left| \bar{B}_{\sss KK}\right|^2 
 \right) = 
 r_u^2 \tilde{\cal T}^2 + \tilde{\cal P}^{\prime 2} 
 - 2 r_u \tilde{\cal T}
 \tilde{\cal P}^\prime
 \cos\delta \cos\gamma, 
\label{D2} \\ 
\tilde{E} &\equiv & \frac{1}{2}
 \left( 
 \left| B_{\sss KK}\right|^2 
 - \left| \bar{B}_{\sss KK}\right|^2 
 \right) = 
 2 r_u \tilde{\cal T}\tilde{\cal P}^\prime
 \sin\delta \sin\gamma, 
\label{E2} \\ 
\tilde{F} &\equiv & \left| B_{\sss KK}^s\right|^2 
 = {\tilde{{\cal P}}^{\prime 2}} 
\label{F2} 
\eea
Combined with the 6 quantities in Eqs.~(\ref{A1}-\ref{F1}), we have 12
equations in 9 unknowns. As shown below, this allows us to solve for the
CP angles, as in the DGR method, but greatly reduces the discrete
ambiguities.

The CP angles can be obtained as follows. First, one finds the ratios
$\tilde{\cal T}/{\cal T}$, $\tilde{\cal P}/{\cal P}$, and ${\tilde{\cal
P}}^\prime/{\cal P}^\prime$:
\beq
a \equiv \frac{\tilde{\cal T}} {\cal T} =
 \frac{\tilde{E}}{E}\sqrt{\frac{F}{\tilde{F}}} ~,~~~~
b \equiv \frac{\tilde{\cal P}}{\cal P} = 
 \frac{\tilde{B}E}{B \tilde{E}}\sqrt{\frac{\tilde{F}}{F}} ~,~~~~
c \equiv \frac{\tilde{\cal P}^\prime}{{\cal P}^\prime} = 
	 \sqrt{\frac{\tilde{F}}{F}} ~.
\label{abc}
\eeq
Using these, we can find the values of all the magnitudes of the
amplitudes. The amplitudes ${\cal T}$ and ${\cal P}$ are obtained from
\beq
{\cal T}^2 = \frac{( a c D - \tilde{D}) - c ( a - c ) F }
 { a ( c - a ) r_u^2 } ~,~~~~
{\cal P}^2 = \frac{a b A - \tilde{A}}{b ( a - b )} + 
 \frac{a}{b} \frac{(a c D - \tilde{D}) - 
 c ( c - a ) F }
 { a ( c - a ) r_u^2} ~,
\label{T2P2}
\eeq
and the remaining amplitudes can be found using Eq.~(\ref{abc}). Note that
all magnitudes are positive, by definition.

We now turn to the angles. Using our knowledge of the magnitudes of the
amplitudes, we have
\bea
\cos(\delta - \alpha ) &=& \frac{ {\cal T}^2 + {\cal P}^2 - A - B }
 { 2 {\cal T P }} ~, \nn \\
\cos(\delta + \alpha ) &=& \frac{ {\cal T}^2 + {\cal P}^2 - A + B }
 { 2 {\cal T P }} ~, \nn \\
\cos(\delta - \gamma ) &=& \frac{ r_u^2 {\cal T}^2 + F - D + E }
 { 2 r_u {\cal T} \sqrt{F}} ~, \nn \\
\cos(\delta + \gamma ) &=& \frac{ r_u^2 {\cal T}^2 + F - D - E }
 { 2 r_u {\cal T} \sqrt{F}} ~.
\label{cosines}
\eea
These equations can be solved to give the phases $\alpha$, $\gamma$ 
and $\delta$ up to a fourfold ambiguity. That is, if $\alpha_0$, $\gamma_0$
and $\delta_0$ are the true values of these phases, then the following
four sets of phases solve the above equations: $\{\alpha_0, \gamma_0,
\delta_0 \}$, $\{-\alpha_0, -\gamma_0, -\delta_0 \}$, $\{\alpha_0 - \pi,
\gamma_0 - \pi, \delta_0 - \pi\}$, and $\{\pi - \alpha_0, \pi - \gamma_0,
\pi - \delta_0 \}$. Note, however, that we still haven't used the $C$ and
$\tilde{C}$ measurements. Their knowledge eliminates two of the four sets,
leaving
\bea
&~& \{\alpha_0, \gamma_0, \delta_0\} ~, \nn \\
&~& \{\alpha_0 - \pi, \gamma_0 - \pi, \delta_0 - \pi\} ~.
\eea
These two solutions correspond to two different orientations of the
unitarity triangle, one pointing up, the other down. This final ambiguity
cannot be resolved by this method alone. However, within the SM it can be
removed by using other measurements such as $\epsilon $ in the kaon system
or the third CP angle $\beta$.

Thus, for the case $\delta \ne 0$, this extension of the DGR method removes
almost all of the discrete ambiguities found in the original method.

\subsection{Special Case: $a=b=c$}

{}From Eq.~(\ref{T2P2}), one can see that the above method will not work
if $a=c$ or $a=b$. We therefore reconsider the analysis in the worst-case
scenario of $a=b=c$. In this case the 12 equations 
(Eqs.~\ref{A1}-\ref{F1}, \ref{A2}-\ref{F2}) reduce to 7:
\bea
A &=& {\cal T}^2 + {\cal P}^2 
 - 2 {\cal T}{\cal P} \cos\delta \cos\alpha~, 
\label{SA1} \\
B &=& - 2 {\cal T}{\cal P} \sin\delta \sin\alpha~, 
\label{SB1} \\
C &=& - {\cal T}^2 \sin2\alpha 
 + 2 {\cal T P} \cos\delta \sin\alpha~, 
\label{SC1} \\
\tilde{C} &=& a^2 \left[ r_u^2 {\cal T}^2 \sin2\gamma 
 - 2 r_u {\cal T }{\cal P}^\prime \cos\delta \sin\gamma \right] ~, 
\label{SC2} \\ 
D &=& r_u^2 {\cal T}^2 + {\cal P}^{\prime 2} 
 - 2 r_u {\cal T}{\cal P}^\prime \cos\delta\cos\gamma~, 
\label{SD1} \\
E &=& 2 r_u {\cal T}{\cal P}^\prime \sin\delta \sin\gamma~, 
\label{SE1} \\ 
F &=& {\cal P}^{\prime 2}~, \label{SF1}
\eea 
with $\tilde{A} = a^2 A$, $\tilde{B} = B$, $\tilde{D} = a^2 D$, $\tilde{E} =
E$, and $\tilde{F} = a^2 F$. The quantity $a$ can therefore be determined by
measurements of $\tilde{A}$ and $A$, for example. Without loss of generality,
we set $a=1$; other values of $a$ correspond simply to different values
of $\tilde{C}$ above.

The original system of 12 equations in 9 unknowns is therefore reduced to
7 equations in 6 unknowns: ${\cal T}$, ${\cal P}$, ${\cal P}^\prime$,
$\alpha$, $\gamma$, and $\delta$. In this case the solution can still be
found without discrete ambiguities, although numerical methods are
required. The key observation is that 6 of the above 7 equations --- 
$A$, $B$, $C$, $D$, $E$, $F$ --- are the same as those of the DGR method in
Section 2 (Eqs.~\ref{A1}-\ref{F1}). These 6 equations can be solved for
the 6 unknowns, up to discrete ambiguities, by following the method of
Ref.~\cite{DR}. However, the spurious solutions can be eliminated since
they do not, in general, satisfy the constraint of the seventh equation,
$\tilde{C}$. Thus, even if $a=b=c$, we can solve for all the parameters, up
to the discrete ambiguity given by $\{\alpha,\gamma,\delta\} \to
\{\alpha-\pi,\gamma-\pi,\delta-\pi\}$, corresponding to the unitarity
triangle pointing up or down.


\section{Unequal Strong Phases in $\Delta S=0$ and $\Delta S = 1$ Decays}

One of the assumptions made by DGR is that the strong phase difference in
the $\Delta S = 0$ sector, $\delta \equiv \delta_{\sss T} - \delta_{\sss
P}$, is the same as that in the $\Delta S = 1$ sector, $\delta' \equiv
\delta_{\sss T'} - \delta_{\sss P'}$. This assumption is necessary in the
DGR method since otherwise there would be 6 equations in 7 unknowns.
However, it is not clear how good this assumption is. $SU(3)$ breaking
could in principle lead to a measurable difference between $\delta$ and
$\delta'$. DGR argue that, since the phases are expected to be small
anyway \cite{Pphase}, this assumption is unlikely to introduce a
significant uncertainty into the method. Still, it does introduce a
possible theoretical error into the procedure.

Fortunately, when one considers in addition $\bs$ decays, this assumption
is no longer necessary. We therefore reconsider the analysis of the
previous section for the case in which $\delta \ne \delta'$. (Note that we
continue to assume that the relative strong phases are independent of the
flavor of the spectator quark, so that there are only two strong phases, 
and not four.) In this case we have 12 equations in 10 unknowns. The 12
equations are very similar to those shown earlier: 
\bea
A &=& {\cal T}^2 + {\cal P}^2 
 - 2 {\cal T}{\cal P} \cos\delta \cos\alpha ~, 
\label{stNA1} \\
\tilde{A} &=& a^2 {\cal T}^2 + b^2 {\cal P}^2 
 - 2 a b {\cal T} {\cal P} \cos\delta \cos\alpha ~, 
\label{stNA2} \\
B &=& - 2 {\cal T}{\cal P} \sin\delta \sin\alpha ~, 
\label{stNB1} \\
\tilde{B} &=& - 2 \tilde{\cal T}\tilde{\cal P} \sin\delta \sin\alpha ~, 
\label{stNB2} \\
C &=& - {\cal T}^2 \sin2\alpha + 2 {\cal T P} \cos\delta \sin\alpha ~, 
\label{stNC1} \\
\tilde{C} &=& a^2 r_u^2 {\cal T}^2 \sin2\gamma 
 - 2 r_u a c {\cal T }{\cal P}^\prime \cos\delta^\prime \sin\gamma ~, 
\label{stNC2} \\ 
D &=& r_u^2 {\cal T}^2 + {\cal P}^{\prime 2} 
 - 2 r_u {\cal T}{\cal P}^\prime \cos\delta^\prime \cos\gamma ~, 
\label{stND1} \\
\tilde{D} &=& a^2 r_u^2 {\cal T}^2 + c^2 {\cal P}^{\prime 2} 
 - 2 r_u a c {\cal T} {\cal P}^\prime \cos\delta^\prime \cos\gamma ~, 
\label{stND2} \\ 
E &=& 2 r_u {\cal T}{\cal P}^\prime \sin\delta^\prime \sin\gamma ~, 
\label{stNE1} \\ 
\tilde{E} &=& 2 r_u \tilde{\cal T}\tilde{\cal P}^\prime \sin\delta
\sin\gamma ~, 
\label{stNE2} \\ 
F &=& {\cal P}^{\prime 2} ~, \label{stNF1} \\
\tilde{F} &=& \tilde{P}^{\prime 2} ~. \label{stNF2}
\eea 

The magnitudes of the amplitudes are found in exactly the same way as in
Section 3 (Eqs.~\ref{abc}-\ref{T2P2}). Following this analysis, we find
\bea
\cos(\delta - \alpha ) &=& \frac{ {\cal T}^2 + {\cal P}^2 - A - B }
 { 2 {\cal T P }} ~,
\label{stCD-A}\\
\cos(\delta + \alpha ) &=& \frac{ {\cal T}^2 + {\cal P}^2 - A + B }
 { 2 {\cal T P }} ~,
\label{stCD+A}\\ 
\cos(\delta^\prime - \gamma ) &=& \frac{ r_u^2 {\cal T}^2 + F - D + E}
 { 2 r_u {\cal T} \sqrt{F}} ~,
\label{stCD-G}\\
\cos(\delta^\prime + \gamma ) &=& \frac{ r_u^2 {\cal T}^2 + F - D - E}
 { 2 r_u {\cal T} \sqrt{F}} ~.
\label{stCD+G} 
\eea
{}From Eqs.~(\ref{stCD-A}) and (\ref{stCD+A}) we can find 4 sets of
solutions for $\{\alpha, \delta\}$. That is, if $\{\alpha_0, \delta_0\}$ are
the true values of these phases, then the equations are solved by the
following sets of phases: $\{\alpha_0, \delta_0 \}$, $\{-\alpha_0, -\delta_0
\}$, $\{\alpha_0 - \pi, \delta_0 - \pi\}$, and $\{\pi - \alpha_0, \pi -
\delta_0 \}$. However, as in Sec.~3, the measurement of $C$ eliminates two
of the sets, leaving $\{\alpha_0, \delta_0 \}$ and $\{\alpha_0 - \pi,
\delta_0 - \pi\}$. Similarly, Eqs.~(\ref{stCD-G}) and (\ref{stCD+G}) give
4 sets of solutions for $\{\gamma, \delta^\prime \}$. But the measurement of
$\tilde{C}$ again eliminates two of them, leaving $\{\gamma_0, \delta'_0
\}$ and $\{\gamma_0 - \pi, \delta'_0 - \pi\}$.

Thus, the measurements in Eqs.~(\ref{stNA1}-\ref{stNF2}) allow the
extraction of the angles $\{\alpha, \delta, \gamma, \delta'\}$ up to a
four-fold ambiguity. However, the definition of the unitarity triangle
requires us to choose the same sign for $\alpha$ and $\gamma$. Thus, once
again, the angles can be extracted up to a two-fold ambiguity:
\bea
&~& \{\alpha_0, \delta_0, \gamma_0, \delta'_0\} ~, \nn \\
&~& \{\alpha_0 - \pi, \delta_0 -\pi, \gamma_0 -\pi, \delta'_0 - \pi\} ~,
\eea
which corresponds to the unitarity triangle pointing up or down.

\subsection{Special Case: $a=b=c$}

As in Sec.~3.1, we consider the special case of $a=b=c$, in which case the
above method breaks down. As before, we choose, without loss of
generality, $a=1$. The 12 equations of Eqs.~(\ref{stNA1}-\ref{stNF2})
reduce to 7:
\bea
A &=& {\cal T}^2 + {\cal P}^2 
	- 2 {\cal T}{\cal P} \cos\delta \cos\alpha~, \label{stSA1} \\
B &=& - 2 {\cal T}{\cal P} \sin\delta \sin\alpha~, \label{stSB1} \\
C &=& - {\cal T}^2 \sin2\alpha 
	+ 2 {\cal T P} \cos\delta \sin\alpha~, \label{stSC1} \\
\tilde{C} &=& r_u^2 {\cal T}^2 \sin2\gamma 
	- 2 r_u {\cal T }{\cal P}^\prime 
	\cos\delta^\prime \sin\gamma~, \label{stSC2} \\ 
D &=& r_u^2 {\cal T}^2 + {\cal P}^{\prime 2} 
	- 2 r_u {\cal T}{\cal P}^\prime
	\cos\delta^\prime \cos\gamma~, \label{stSD1} \\
E &=& 2 r_u {\cal T}{\cal P}^\prime
	\sin\delta^\prime \sin\gamma~, \label{stSE1} \\ 
F &=& {\cal P}^{\prime 2}~. \label{stSF1}
\eea 
Since there are 7 unknowns, these 7 equations can be solved, but there
will be discrete ambiguities.

The solutions can be obtained as follows. First, from Eqs.~(\ref{stSA1}),
(\ref{stSB1}), (\ref{stSD1}), and (\ref{stSE1}), we have
\bea 
\cos\alpha &=& \frac{{\cal T}^2 + {\cal P }^2 - A }
	{ 2 {\cal T P} \cos\delta }~, \label{stScosa}\\
\sin\alpha &=& - \frac{B}{2 {\cal T P} \sin\delta }~, \label{stSsina}\\
\cos\gamma &=& \frac{r_u^2 {\cal T}^2 + F - D }
	{ 2 r_u {\cal T} \sqrt{F} \cos\delta^\prime }~,\label{stScosg}\\
\sin\gamma &=& \frac{E}{2 r_u {\cal T}\sqrt{F} \sin\delta^\prime }~. 
\label{stSsing}
\eea
The angles $\alpha$ and $\gamma$ can be eliminated from the above equations,
yielding quadratic equations for $\cos^2\delta$ and $\cos^2\delta^\prime$:
\bea
4 {\cal T}^2 {\cal P}^2 \cos^4\delta - \left\{ 4 {\cal T}^2 {\cal P}^2 + 
( {\cal T}^2 + {\cal P}^2 - A )^2 - B^2 \right\} \cos^2\delta 
~~~~~ & ~ & \nn \\
+ ~( {\cal T}^2 + {\cal P}^2 - A )^2 & = & 0 ~, 
\label{stSCD41} \\
- 4 r_u^2 {\cal T}^2 F \cos^4\delta^\prime+\left\{ 4 r_u^2 {\cal T}^2 F-
E^2 + ( r_u^2 {\cal T}^2 + F - D)^2 \right\} \cos^2\delta^\prime 
~~~~~ & ~ & \nn \\
- ~( r_u^2 {\cal T}^2 + F - D)^2 & = & 0 ~.
\label{stTCDP2}
\eea
Eqs.~(\ref{stSC1}), (\ref{stScosa}) and (\ref{stSsina}) can be combined to
yield another quadratic equation in $\cos^2\delta$. Similarly,
Eqs.~(\ref{stSC2}), (\ref{stScosg}) and (\ref{stSsing}) combine to give
another quadratic equation in $\cos^2\delta^\prime$. These new equations
are:
\bea
4 {\cal P}^4 \left( C^2 + B^2 \right) \cos^4\delta 
- \left\{ 4 {\cal P}^4 C^2 + 4 {\cal P}^2 B^2 ( {\cal T}^2+{\cal P}^2-
A) \right\} \cos^2\delta 
~~~~~ & ~ & \nn \\
+ ~B^2 ( {\cal T}^2 + {\cal P}^2 - A)^2 & = & 0 ~,
\label{stSCD42} \\
4 F^2 \left( E^2 + \tilde{C}^2 \right) 
\cos^4\delta^\prime - \left\{ 4 F E^2 ( r_u^2
{\cal T}^2 + F - D) + 4 F^2 \tilde{C}^2 \right\} \cos^2\delta^\prime 
~~~~~ & ~ & \nn \\
+ ~E^2 ( r_u^2 {\cal T}^2 + F - D )^2 & = & 0 ~.
\label{stTCDP1}
\eea
Note that Eqs.~(\ref{stSCD42}) and (\ref{stTCDP1}) depend on $C^2$
and $\tilde{C}^2$, so that some sign information has been (temporarily)
lost. Eqs.~(\ref{stSCD41}-\ref{stTCDP1}) can now be solved
straightforwardly to give ${\cal T}$, ${\cal P}$, $\cos^2\delta$
and $\cos^2\delta^\prime$, and the CP angles $\alpha$ and $\gamma$ can be
obtained. Of course, since the equations are quadratic, there are
multiple solutions for all these quantities. Some of these solutions can
be eliminated by now reconsidering Eqs.~(\ref{stSC1}) and (\ref{stSC2}),
{\it i.e.} the signs of $C$ and $\tilde{C}$. Still, many solutions remain.

\begin{table}
{\scriptsize
\begin{center}
\begin{tabular}{|c c c c|c|c|c|c|c|c|c|}
\hline 
$\alpha_{in}$ &$\gamma_{in}$ &$\delta_{in}$ &$\delta_{in}^\prime$
 &$\alpha_{out}$ 
 &$\gamma_{out}$ &$\delta_{out}$ &$\delta_{out}^\prime$ 
 &${\cal T}_{out}$ &${\cal P}_{out}$ & Notes \\ \hline
45.0 & 120.0 & 84.3 & 36.9 & 47.3 & 120.0
 & 84.8 & 36.9 & 1.00 & 0.27 & c \\ 
\ & \ & \ & \ & 45.0 & 120.0
 & 84.3 & 36.9 & 1.00 & 0.28 & a \\ 
\ & \ & \ & \ & 132.7 & 120.0
 & 168.6 & 36.9 & 1.00 & 1.36 & b \\
\ & \ & \ & \ & 135.0 & 120.0
 & 168.6 & 36.9 & 1.00 & 1.41 & b \\
\ & \ & \ & \ & 8.4 & 54.8 
 & 3.9 & 7.5 & 4.86 & 4.14 & b \\
\ & \ & \ & \ & 8.7 & 54.8
 & 3.7 & 7.5 & 4.86 & 4.16 & b \\
\ & \ & \ & \ & 171.3 & 54.8
 & 177.2 & 7.5 & 4.86 & 5.47 & b \\
\ & \ & \ & \ & 171.6 & 54.8
 & 177.1 & 7.5 & 4.86 & 5.50 & b \\
\ & \ & \ & \ & 7.0 & 137.0
 & 3.1 & 172.5 & 5.83 & 5.11 & b \\ 
\ & \ & \ & \ & 7.2 & 137.0
 & 3.0 & 172.5 & 5.83 & 5.14 & b \\ 
\ & \ & \ & \ & 172.8 & 137.0 
 & 177.6 & 172.5 & 5.83 & 6.44 & b \\ 
\ & \ & \ & \ & 173.0 & 137.0 
 & 177.5 & 172.5 & 5.83 & 6.47 & b \\ 
\ & \ & \ & \ & 6.5 & 172.0 
 & 2.9 & 143.1 & 6.26 & 5.54 & b \\ 
\ & \ & \ & \ & 6.7 & 172.0
 & 2.8 & 143.1 & 6.26 & 5.57 & b \\
\ & \ & \ & \ & 173.3 & 172.0 
 & 177.8 & 143.1 & 6.26 & 6.88 & b \\ 
\ & \ & \ & \ & 173.5 & 172.0
 & 177.7 & 143.1 & 6.26 & 6.90 & b \\ 
\hline
70.0 & 90.0 & 84.3 & 36.9 & 70.0 & 90.0
 & 84.3 & 36.9 & 1.00 & 0.24 & a \\
\ & \ & \ & \ & 23.6 & 90.0 
 & 73.4 & 36.9 & 1.00 & 0.60 & b \\ 
\ & \ & \ & \ & 130.0 & 90.0
 & 159.7 & 36.9 & 1.00 & 0.70 & b \\
\ & \ & \ & \ & 156.4 & 90.0 
 & 160.0 & 36.9 & 1.00 & 1.76 & b \\
\ & \ & \ & \ & 5.1 & 52.7 
 & 8.7 & 9.7 & 4.48 & 3.76 & b \\
\ & \ & \ & \ & 12.1 & 52.7 
 & 3.4 & 9.7 & 4.48 & 4.07 & b \\
\ & \ & \ & \ & 167.9 & 52.7
 & 177.0 & 9.7 & 4.48 & 4.70 & b \\
\ & \ & \ & \ & 174.9 & 52.7 
 & 173.7 & 9.7 & 4.48 & 5.24 & b \\
\ & \ & \ & \ & 4.1 & 140.4
 & 6.7 & 170.3 & 5.59 & 4.87 & b \\
\ & \ & \ & \ & 9.7 & 140.4
 & 2.7 & 170.3 & 5.59 & 5.20 & b \\
\ & \ & \ & \ & 170.3 & 140.4
 & 177.6 & 170.3 & 5.59 & 5.84 & b \\
\ & \ & \ & \ & 175.9 & 140.4 
 & 174.8 & 170.3 & 5.59 & 6.35 & b \\
\ & \ & \ & \ & 4.0 & 170.0 
 & 6.5 & 143.1 & 5.79 & 5.06 & b \\
\ & \ & \ & \ & 9.3 & 170.0
 & 2.6 & 143.1 & 5.79 & 5.40 & b \\
\ & \ & \ & \ & 170.7 & 170.0
 & 177.7 & 143.1 & 5.79 & 6.03 & b \\
\ & \ & \ & \ & 176.0 & 170.0 
 & 175.0 & 143.1 & 5.79 & 6.54 & b \\
\hline
100.0 & 60.0 & 84.3 & 36.9 & 100.0 & 60.0
 & 84.3 & 36.9 & 1.00 & 0.20 & a \\
\ & \ & \ & \ & 80.0 & 60.0
 & 28.7 & 36.9 & 1.00 & 0.42 & c \\
\ & \ & \ & \ & 163.8 & 60.0 
 & 131.3 & 36.9 & 1.00 & 0.76 & b \\
\ & \ & \ & \ & 16.2 & 60.0
 & 23.4 & 36.9 & 1.00 & 1.79 & b \\
\ & \ & \ & \ & 176.3 & 47.9 
 & 168.8 & 9.4 & 4.29 & 3.66 & b \\
\ & \ & \ & \ & 166.7 & 47.9 
 & 177.1 & 9.4 & 4.29 & 3.98 & b \\
\ & \ & \ & \ & 13.3 & 47.9 
 & 2.6 & 9.4 & 4.29 & 4.37 & b \\
\ & \ & \ & \ & 3.7 & 47.9 
 & 8.1 & 9.4 & 4.29 & 5.01 & b \\
\ & \ & \ & \ & 176.9 & 142.6
 & 171.1 & 170.6 & 5.23 & 4.60 & b \\
\ & \ & \ & \ & 169.2 & 142.6 
 & 177.8 & 170.6 & 5.23 & 4.95 & b \\
\ & \ & \ & \ & 10.8 & 142.6
 & 2.3 & 170.6 & 5.23 & 5.34 & b \\
\ & \ & \ & \ & 3.1 & 142.6
 & 6.8 & 170.6 & 5.23 & 5.95 & b \\
\ & \ & \ & \ & 177.0 & 170.5 
 & 171.2 & 143.1 & 5.27 & 4.64 & b \\
\ & \ & \ & \ & 169.2 & 170.5 
 & 177.7 & 143.1 & 5.27 & 4.99 & b \\
\ & \ & \ & \ & 10.8 & 170.5 
 & 2.1 & 143.1 & 5.27 & 5.38 & b \\
\ & \ & \ & \ & 3.0 & 170.5 
 & 6.8 & 143.1 & 5.27 & 5.99 & b \\
\hline
\end{tabular}
\end{center} }
\caption{Output values of the strong and weak phases, as well as the
amplitudes, for given values of the input strong phases, and weak phases,
$\alpha_{in}$ and $\gamma_{in}$. All phase angles are given in degrees. In
the `Notes' column, `a', `b' and `c' indicate respectively the correct
solution, a solution inconsistent with other experimental constraints, and
a potential ambiguity.}
\label{Sec4-Tab1}
\end{table}

In Table \ref{Sec4-Tab1}, we show examples of some of these solutions for
various values of the parameters. For the amplitudes, we take ${\cal T } =
\tilde{\cal T} = 1$, ${\cal P}^\prime = \tilde{\cal P}^\prime= 1$, and 
${\cal P} = \tilde{\cal P} = \frac{|V_{us}|}{|V_{ud}|} 
\frac{\sin\gamma}{\sin\alpha } {\cal P}^\prime$. The values assumed for the
weak and strong phases are shown in the table.

This table shows that there are indeed many solutions for these equations.
(And note that there are additional solutions which we have not listed, in
which the phases $\{\alpha_{out}, \gamma_{out}, \delta_{out},
\delta'_{out}\}$ are changed to $\{\alpha_{out}-\pi, \gamma_{out}-\pi,
\delta_{out}-\pi, \delta'_{out}-\pi\}$.) However, not all solutions are
allowed within the context of the SM. For example, present experimental
information constrains $20^\circ \lsim \alpha \lsim 120^\circ$, $30^\circ
\lsim \gamma \lsim 150^\circ$, and  $10^\circ \lsim \beta \lsim 45^\circ$
\cite{AliLon}. Following Refs.~\cite{DGR,DR}, in the table we have labeled
the solutions as follows: (a) correct solution, (b) one or more of the CP
angles outside of the SM domain, and (c) potential ambiguity. As is clear
from the table, most of the spurious solutions disappear when one imposes
the SM constraints. Of course, one might be overlooking the presence of new
physics in this way, but that is one of the problems caused by the presence
of discrete ambiguities.


\section{Non-negligible $u$- and $c$-quark Penguin Contributions}

Another assumption made by DGR is that the $b \to d$ penguin is dominated
by internal $t$-quarks. This has the effect that the weak phase of the
penguin is simply $-\beta$. However, Buras and Fleischer have argued that
the contributions of internal $u$- and $c$-quarks to $b \to d$ penguins are
not negligible \cite{Fleischer,BF}. They estimate that these additional
diagrams can be between 20\% and 50\% of the leading $t$-quark contribution.
By their own admission, this is a very rough estimate, but it indicates a
potential complication to the method of DGR, as well as its extension.

If their estimate is correct, these additional contributions must be taken
into account. The simplest way to do this is to write:
\beq
P = \sum_{q=u,c,t} V_{qd}^* V_{qb} \, P_q = V_{ud}^* V_{ub} \, 
(P_u - P_c) + V_{td}^* V_{tb} \, (P_t - P_c),
\label{penguins}
\eeq
where the unitarity of the CKM matrix has been used. The key point here is
that the combination of CKM matrix elements $V_{ud}^* V_{ub}$, which
multiplies the new contributions, is the same as that which appears in the
tree diagram. Thus, the amplitude for $B^0 \to \pi\pi$ can be written
\bea
A_{\pi\pi} & = & {\cal T} e^{i \delta_T} e^{i\gamma } + 
( {\cal P}_u e^{i\delta_u} - {\cal P}_c e^{i\delta_c} ) e^{i\gamma } +
( {\cal P}_t e^{i\delta_t} - {\cal P}_c e^{i\delta_c} ) e^{- i\beta }
\nn \\ 
& \equiv & {\cal T}_{\sss P} e^{i \delta_{T_P}} e^{i\gamma } +
{\cal P}_{\sss P} e^{i\delta_{P_P}} e^{- i\beta } ~,
\eea
where ${\cal T}_{\sss P}$ and $\delta_{\sss T_P}$ include the tree and $u$-
and $c$-quark penguin pieces, and similarly for ${\cal P}_{\sss P}$ and
$\delta_{\sss P_P}$. Likewise, the amplitude for $\bs\rightarrow \pi^+
K^-$ can be written
\bea
B_{\sss \pi K} & = & \tilde{{\cal T}} e^{i \delta_T} e^{i\gamma } +
( {\tilde{\cal P}}_u e^{i\delta_u} - {\tilde{\cal P}}_c e^{i\delta_c} )
e^{i\gamma } + 
( {\tilde{\cal P}}_t e^{i\delta_t} - {\tilde{\cal P}}_c e^{i\delta_c} )
e^{- i\beta } \nn \\
& \equiv & {\tilde{\cal T}}_{\sss P} e^{i \delta_{T_P}} e^{i\gamma } +
{\tilde{\cal P}}_{\sss P} e^{i\delta_{P_P}} e^{- i\beta } ~.
\eea
Note that the new penguin contributions modify the sizes and phases of the
tree and penguin amplitudes, but leave the forms of $A_{\pi\pi}$ and
$B_{\sss \pi K}$ unchanged.

The $b\to s$ penguin is not affected in the same way. For this penguin, one
can perform a similar decomposition as in Eq.~(\ref{penguins}). However,
in this case, the contribution proportional to $P_t - P_c$ dominates, since
$|V_{us}^* V_{ub}| \ll |V_{ts}^* V_{tb}|$. Thus, the remaining amplitudes
can be written 
\bea
A_{\sss \pi K} & = & {\cal T}' e^{i \delta_T} e^{i\gamma } -
 {\cal P}^\prime_{\sss P} e^{i \delta_{P_P}} ~, \\
B_{\sss KK} & = & \tilde{{\cal T}}' e^{i \delta_T} e^{i\gamma } -
 \tilde{{\cal P}}^\prime_{\sss P} e^{i \delta_{P_P}} ~, \\
A_{\sss \pi K}^+ & = & {\cal P}^\prime_{\sss P} e^{i \delta_{P_P}}~,\\
B_{\sss KK}^s &=& \tilde{{\cal P}}^\prime_{\sss P} e^{i \delta_{P_P}}~.
\eea
Looking at the above 6 amplitudes, there are two points to be noted. 
First, the inclusion of the $u$- and $c$-quark penguins modifies the strong
phase appearing in the $\Delta S = 0$ processes: $\delta = \delta_{\sss
T_P} - \delta_{\sss P_P}$. This is clearly not the same as that appearing
in $\Delta S = 1$ processes: $\delta' = \delta_{\sss T} - \delta_{\sss
P_P}$. Thus, consideration of $u$- and $c$-quark penguins requires us to take
$\delta \ne \delta'$. Second, the presence of these additional penguin
contributions destroys the relation between the tree contributions of
$\Delta S = 0$ and $\Delta S = 1$ processes: although ${\cal T}'/{\cal T} =
r_u$, ${\cal T}'/{\cal T}_{\sss P} \ne r_u$. Therefore ${\cal T}'$ must be
left as an independent parameter. We will return to this point below.

With the above modified amplitudes, the 12 measurements are
\bea
A &=& {\cal T}_{\sss P}^2 + {\cal P}_{\sss P}^2 - 
2 {\cal T}_{\sss P}{\cal P}_{\sss P} \cos\delta \cos\alpha, 
\label{GA1} \\
\tilde{A} &=& \tilde{\cal T}_{\sss P}^2 + \tilde{\cal P}_{\sss P}^2 
- 2 \tilde{\cal T}_{\sss P} 
\tilde{\cal P}_{\sss P} \cos\delta \cos\alpha, 
\label{GA2} \\
B &=& - 2 {\cal T}_{\sss P}{\cal P}_{\sss P} \sin\delta \sin\alpha, 
\label{GB1} \\
\tilde{B} &=&-2\tilde{\cal T}_{\sss P}\tilde{\cal P}_{\sss P}\sin\delta
\sin\alpha, 
\label{GB2} \\
C &=& - {\cal T}_{\sss P}^2 \sin2\alpha + 2 {\cal T}_{\sss P} 
{\cal P}_{\sss P} \cos\delta \sin\alpha, 
\label{GC1} \\
\tilde{C} &=& \tilde{\cal T}^{\prime 2} \sin2\gamma - 2
\tilde{\cal T}^\prime \tilde{\cal P}_{\sss P}^\prime
\cos\delta^\prime \sin\gamma,
\label{GC2} \\ 
D &=& {\cal T}^{\prime 2} + {\cal P}_{\sss P}^{\prime 2} 
-2{\cal T}^\prime {\cal P}_{\sss P}^\prime\cos\delta^\prime \cos\gamma, 
\label{GD1} \\
\tilde{D} &=& \tilde{\cal T}^{\prime 2} + 
\tilde{\cal P}_{\sss P}^{\prime 2} - 2 \tilde{\cal T}^\prime 
\tilde{\cal P}_{\sss P}^\prime
\cos\delta^\prime \cos\gamma, 
\label{GD2} \\ 
E &=& 2 {\cal T}^\prime {\cal P}_{\sss P}^\prime \sin\delta^\prime
\sin\gamma, 
\label{GE1} \\ 
\tilde{E} &=& 2 \tilde{\cal T}^\prime \tilde{\cal P}_{\sss P}^\prime
\sin\delta^\prime \sin\gamma, 
\label{GE2} \\ 
F &=& P_{\sss P}^{\prime 2}, 
\label{GF1} \\
\tilde{F} &=& \tilde{P}_{\sss P}^{\prime 2}, 
\label{GF2} 
\eea
where $\delta = \delta_{\sss T_P} - \delta_{\sss P_P}$ and $\delta' =
\delta_{\sss T} - \delta_{\sss P_P}$. Thus we end up with 12 equations in
12 unknowns. 

However, there is a problem. An examination of the above equations reveals
that the 12 measurements separate into two independent categories, those
for $\Delta S = 0$ processes, and those for $\Delta S = 1$. The 5
measurements in the $\Delta S = 0$ sector, Eqs.~(\ref{GA1}-\ref{GC1}),
depend only on the 6 parameters${\cal T}_{\sss P}$, $\tilde{\cal T}_{\sss
P}$, ${\cal P}_{\sss P}$, $\tilde{\cal P}_{\sss P}$, $\alpha$, and $\delta$. 
Since the $\Delta S = 0$ sector has 5 equations in 6 unknowns, it is
therefore impossible to extract the CP angle $\alpha$. Thus, the DGR method,
as well as its extension, breaks down when the $u$- and $c$-quark penguin
contributions are included.

Before discussing the $\Delta S = 1$ processes, let us examine why the
method breaks down in this case. The crucial problem is that, in the
presence of the additional penguin contributions, the relation ${\cal
T}'/{\cal T} = r_u$, which takes into account $SU(3)$ breaking, is
apparently no longer valid. However, it is not clear how badly this
relation is violated. Including the $u$- and $c$-quark penguin
contributions, we have
\bea
{{\cal T}' \over {\cal T}_{\sss P}} &=&{ {\cal T}' e^{i \delta_T} \over 
{\cal T} e^{i \delta_T} + 
{\cal P}_u e^{i\delta_u} - {\cal P}_c e^{i\delta_c} } \nn \\
& \simeq & r_u \left[ 1 - { 
{\cal P}_u e^{i\delta_u} - {\cal P}_c e^{i\delta_c} \over 
{\cal T} e^{i \delta_T} } \right] .
\eea
Buras and Fleischer found the ratio$|(P_c - P_u)/(P_t - P_u)|$ to be
between 20\% and 50\% \cite{Fleischer,BF}. However, this does not give us
any information about the ratio$|(P_u - P_c)/T|$. Even if the $u$- and
$c$-quark penguin contributions are sizeable when compared to the $t$-quark
penguin, it may still be that they are quite a bit smaller than the tree
diagram. In this case, we would still have ${\cal T}'/{\cal T}_{\sss P}
\simeq r_u$, and the situation would reduce to that of the previous
section, with 12 equations in 10 unknowns. As shown in that section, the
CP angles $\alpha$ and $\gamma$ can both be found, up to a 2-fold
ambiguity. Thus, even taking into account the $u$- and $c$-quark penguins, 
it may be possible to extract $\alpha$. However, it is difficult to know
for sure, and this will introduce some theoretical uncertainty into the
method.

We now turn to the 7 measurements in $\Delta S = 1$ processes,
Eqs.~(\ref{GC2}-\ref{GF2}), and assume that the $u$- and $c$-quark penguins
are sizeable (since otherwise the method of the previous section holds). 
In this case these measurements depend only on the remaining 6 parameters
${\cal T}'$, $\tilde{\cal T}^\prime$, ${\cal P}_{\sss P}'$, $\tilde{\cal
P}_{\sss P}^\prime$, $\gamma$, and $\delta'$. As we have shown in previous
sections, this implies that $\gamma$ can be extracted up to a 2-fold
ambiguity. 

The solution can be explicitly constructed as follows. As before, we
write $\tilde{\cal T}' = a {\cal T}'$, where $a$ is defined in
Eq.~(\ref{abc}). Using Eqs.~(\ref{GD1}) and (\ref{GD2}), we can then solve
for ${\cal T}'$: 
\beq {\cal T}' = \sqrt{ { \sqrt{\tilde F} 
( D - F ) - \sqrt{F} ( {\tilde D}
- {\tilde F} ) \over \sqrt{\tilde F} - a^2 \sqrt{F} } } ~.
\eeq
Given values for all the amplitudes, Eqs.~(\ref{GD1}) and (\ref{GE1}) can
then be used to obtain the phases $\gamma$ and $\delta'$, up to a 4-fold
ambiguity. Finally, Eq.~(\ref{GC2}) can be used to eliminate two of these
solutions.

This is interesting in its own right, as it is a new method for
extracting $\gamma$. Note that the $\Delta S = 0$ processes are not needed
at all. The four decays which need to be measured here are $\bd \to \pi^-
K^+$, $B^+ \to \pi^+ K^0$, $\bs(t) \to K^+ K^-$, and $\bs \to K^0 {\bar
K^0}$. The assumption of $SU(3)$ symmetry is rather minimal here -- one
assumes only that the strong phases are independent of the spectator
quarks. (Even this assumption is relaxed in the next section.) In
essence, this method removes the penguin contribution from the CP-violating
asymmetry in $\bs(t) \to K^+ K^-$. We will have more to say about this
method in the next section.


\section{Different Strong Phases for Different Spectator Quarks}

The one assumption which we have continued to make throughout the previous
sections is that the strong dynamics ({\it i.e.} the strong phases) is
independent of the flavor of the spectator quark. Given the success of the
spectator model in $B$ decays, this is probably justified. Nevertheless, in
this section we explore the consequences of relaxing this assumption.

If the flavor $SU(3)$ symmetry were unbroken, there would be a single
strong phase, $\delta$. In the decays considered in this paper, there are
two distinct ways in which $SU(3)$ can be broken: (i) $\Delta S=0$ processes
{\it vs.} $\Delta S=1$ processes, and (ii) $\bd$ decays {\it vs.} $\bs$
decays. Therefore in the general case one must consider four different
strong phases:
\bea
\delta & : & (\bd~{\rm or}~B^+~{\rm decays}, \Delta S = 0), \nn \\
\delta' & : & (\bd~{\rm or}~B^+~{\rm decays}, \Delta S = 1), \nn \\
\delta_s & : & (\bs~{\rm decays}, \Delta S = 0), \nn \\
\delta'_s & : & (\bs~{\rm decays}, \Delta S = 1).
\eea
We will assume, however, that the $u$- and $c$-quark penguins are
unimportant. In Section 5, it was shown that if these contributions are
sizeable when compared to the tree diagram the angle $\alpha$ cannot be
extracted. Since in this section we are adding more parameters, the
situation is even worse, and the only way that any information can be
obtained is if the $u$- and $c$-quark penguins are in fact negligible
compared to the tree diagram.

In this case, measurements of the various processes yield the following 12
quantities: 
\bea
A & = & {\cal T}^2 + {\cal P}^2 - 2 {\cal T}{\cal P} \cos\delta
\cos\alpha, 
\label{DsA1} \\
\tilde{A} & = & \tilde{\cal T}^2 + \tilde{\cal P}^2 
	- 2 \tilde{\cal T} \tilde{\cal P} \cos\delta_s \cos\alpha, 
\label{DsA2} \\
B & = & - 2 {\cal T}{\cal P} \sin\delta \sin\alpha, 
\label{DsB1} \\
\tilde{B} &=&-2\tilde{\cal T}\tilde{\cal P} \sin\delta_s \sin\alpha, 
\label{DsB2} \\
C & = & - {\cal T}^2 \sin2\alpha + 2 {\cal T P} \cos\delta \sin\alpha, 
\label{DsC1} \\
\tilde{C} & = & r_u^2 \tilde{\cal T}^2 \sin2\gamma 
 - 2 r_u \tilde{\cal T }\tilde{\cal P}^\prime \cos\delta'_s \sin\gamma, 
\label{DsC2} \\ 
D & = & r_u^2 {\cal T}^2 + {\cal P}^{\prime 2} 
	- 2 r_u {\cal T}{\cal P}^\prime \cos\delta' \cos\gamma, 
\label{DsD1} \\
\tilde{D} & = & r_u^2 \tilde{\cal T}^2 + \tilde{\cal P}^{\prime 2} 
- 2 r_u \tilde{\cal T} \tilde{\cal P}^\prime \cos\delta'_s \cos\gamma, 
\label{DsD2} \\ 
E & = & 2 r_u {\cal T}{\cal P}^\prime \sin\delta' \sin\gamma, 
\label{DsE1} \\ 
\tilde{E} & = & 2 r_u \tilde{\cal T}\tilde{\cal P}^\prime \sin\delta'_s
\sin\gamma, 
\label{DsE2} \\ 
F & = & {\cal P}^{\prime 2} 
\label{DsF1} \\
\tilde{F} & = & \tilde{\cal P}^{\prime 2} 
\label{DsF2} 
\eea
Since there are 12 equations in 12 unknowns, this system of equations can
be solved, but there will be discrete ambiguities.

\begin{table}
{\scriptsize
\begin{center}
\vspace{-0.7cm}
\begin{tabular}{|c|c|c|c|c|c|c|c|c|c|c|}
\hline 
$\alpha_{out}$ &$\gamma_{out}$ &$\delta_{out}$ &$\delta^\prime_{out}$
 &$\delta_{s,out}$ &$\delta^\prime_{s,out}$ 
 &${\cal T}_{out}$&$\tilde{\cal T}_{out}$ 
 &${\cal P}_{out}$ &$\tilde{\cal P}_{out}$ & Notes \\ \hline
132.7 & 120.0 & 168.6 & 5.7 & 5.3
 & 36.9 & 1.00 & 0.80 & 1.36 & 0.32 & b \\
132.7 & 120.0 & 168.6 & 5.7 & 90.0
 & 36.9 & 1.00 & 0.80 & 1.36 & 0.67 & b \\
132.7 & 120.0 & 168.6 & 5.7 & 178.8
 & 36.9 & 1.00 & 0.80 & 1.36 & 1.40 & b \\
135.0 & 120.0 & 168.6 & 5.7 & 5.7
 & 36.9 & 1.00 & 0.80 & 1.41 & 0.31 & b \\
135.0 & 120.0 & 168.6 & 5.7 & 90.0
 & 36.9 & 1.00 & 0.80 & 1.41 & 0.67 & b \\
135.0 & 120.0 & 168.6 & 5.7 & 178.8
 & 36.9 & 1.00 & 0.80 & 1.41 & 1.44 & b \\
47.3 & 120.0 & 84.8 & 5.7 & 173.8
 & 36.9 & 1.00 & 0.80 & 0.27 & 0.32 & c \\
47.3 & 120.0 & 84.8 & 5.7 & 90.0
 & 36.9 & 1.00 & 0.80 & 0.27 & 0.67 & c \\
47.3 & 120.0 & 84.8 & 5.7 & 1.2
 & 36.9 & 1.00 & 0.80 & 0.27 & 1.40 & c \\
45.0 & 120.0 & 84.3 & 5.7 & 174.3
 & 36.9 & 1.00 & 0.80 & 0.28 & 0.31 & a \\
45.0 & 120.0 & 84.3 & 5.7 & 90.0
 & 36.9 & 1.00 & 0.80 & 0.28 & 0.67 & a$^\prime$ \\
45.0 & 120.0 & 84.3 & 5.7 & 1.2
 & 36.9 & 1.00 & 0.80 & 0.28 & 1.44 & a$^\prime$ \\
19.4 & 120.0 & 11.9 & 90.0 & 165.3
 & 36.9 & 2.13 & 0.80 & 1.36 & 0.26 & b \\
19.4 & 120.0 & 11.9 & 90.0 & 90.0
 & 36.9 & 2.13 & 0.80 & 1.36 & 0.67 & b \\
19.4 & 120.0 & 11.9 & 90.0 & 2.1
 & 36.9 & 2.13 & 0.80 & 1.36 & 1.76 & b \\
20.2 & 120.0 & 11.3 & 90.0 & 165.9
 & 36.9 & 2.13 & 0.80 & 1.36 & 0.26 & c \\
20.2 & 120.0 & 11.3 & 90.0 & 90.0
 & 36.9 & 2.13 & 0.80 & 1.36 & 0.67 & c \\
20.2 & 120.0 & 11.3 & 90.0 & 2.1
 & 36.9 & 2.13 & 0.80 & 1.36 & 1.75 & c \\
159.8 & 120.0 & 174.2 & 90.0 & 14.1
 & 36.9 & 2.13 & 0.80 & 2.67 & 0.26 & b \\
159.8 & 120.0 & 174.2 & 90.0 & 90.0
 & 36.9 & 2.13 & 0.80 & 2.67 & 0.67 & b \\
159.8 & 120.0 & 174.2 & 90.0 & 177.9
 & 36.9 & 2.13 & 0.80 & 2.67 & 1.75 & b \\
160.6 & 120.0 & 174.1 & 90.0 & 14.7
 & 36.9 & 2.13 & 0.80 & 2.70 & 0.26 & b \\
160.6 & 120.0 & 174.1 & 90.0 & 90.0
 & 36.9 & 2.13 & 0.80 & 2.70 & 0.67 & b \\
160.6 & 120.0 & 174.1 & 90.0 & 177.9
 & 36.9 & 2.13 & 0.80 & 2.70 & 1.76 & b \\
8.9 & 120.0 & 4.2 & 178.8 & 149.1
 & 36.9 & 4.56 & 0.80 & 3.84 & 0.27 & b \\
8.9 & 120.0 & 4.2 & 178.8 & 90.0
 & 36.9 & 4.56 & 0.80 & 3.84 & 0.67 & b \\
8.9 & 120.0 & 4.2 & 178.8 & 4.4
 & 36.9 & 4.56 & 0.80 & 3.84 & 1.82 & b \\
9.3 & 120.0 & 4.0 & 178.8 & 150.1
 & 36.9 & 4.56 & 0.80 & 3.86 & 0.27 & b \\
9.3 & 120.0 & 4.0 & 178.8 & 90.0
 & 36.9 & 4.56 & 0.80 & 3.86 & 0.67 & b \\
9.3 & 120.0 & 4.0 & 178.8 & 4.3
 & 36.9 & 4.56 & 0.80 & 3.86 & 1.82 & b \\
170.7 & 120.0 & 177.0 & 178.8 & 29.9
 & 36.9 & 4.56 & 0.80 & 5.16 & 0.27 & b \\
170.7 & 120.0 & 177.0 & 178.8 & 90.0
 & 36.9 & 4.56 & 0.80 & 5.16 & 0.67 & b \\
170.7 & 120.0 & 177.0 & 178.8 & 175.7
 & 36.9 & 4.56 & 0.80 & 5.16 & 1.82 & b \\
171.1 & 120.0 & 176.9 & 178.8 & 30.9
 & 36.9 & 4.56 & 0.80 & 5.19 & 0.27 & b \\
171.1 & 120.0 & 176.9 & 178.8 & 90.0
 & 36.9 & 4.56 & 0.80 & 5.19 & 0.67 & b \\
171.1 & 120.0 & 176.9 & 178.8 & 175.6
 & 36.9 & 4.56 & 0.80 & 5.19 & 1.82 & b \\
7.9 & 53.6 & 3.7 & 1.2 & 0.5
 & 6.8 & 5.11 & 4.39 & 4.39 & 3.50 & b \\
7.9 & 53.6 & 3.7 & 1.2 & 0.3
 & 6.8 & 5.11 & 4.39 & 4.39 & 5.19 & b \\
8.3 & 53.6 & 3.5 & 1.2 & 2.3
 & 6.8 & 5.11 & 4.39 & 4.42 & 3.52 & b \\
8.3 & 53.6 & 3.5 & 1.2 & 2.3
 & 6.8 & 5.11 & 4.39 & 4.42 & 5.17 & b \\
171.7 & 53.6 & 177.3 & 1.2 & 177.7
 & 6.8 & 5.11 & 4.39 & 5.72 & 3.52 & b \\
171.7 & 53.6 & 177.3 & 1.2 & 177.7
 & 6.8 & 5.11 & 4.39 & 5.72 & 5.17 & b \\
172.1 & 53.6 & 177.2 & 1.2 & 179.5
 & 6.8 & 5.11 & 4.39 & 5.75 & 3.50 & b \\
172.1 & 53.6 & 177.2 & 1.2 & 179.7
 & 6.8 & 5.11 & 4.39 & 5.75 & 5.19 & b \\
6.8 & 136.8 & 3.1 & 178.8 & 0.4
 & 173.2 & 5.99 & 5.17 & 5.24 & 4.29 & b \\
6.8 & 136.8 & 3.1 & 178.8 & 0.3
 & 173.2 & 5.99 & 5.17 & 5.24 & 5.97 & b \\
7.1 & 136.8 & 2.9 & 178.8 & 0.4
 & 173.2 & 5.99 & 5.17 & 5.27 & 4.30 & b \\
7.1 & 136.8 & 2.9 & 178.8 & 0.3
 & 173.2 & 5.99 & 5.17 & 5.27 & 5.95 & b \\
172.9 & 136.8 & 177.6 & 178.8 & 179.6
 & 173.2 & 5.99 & 5.17 & 6.57 & 4.30 & b \\
172.9 & 136.8 & 177.6 & 178.8 & 179.7
 & 173.2 & 5.99 & 5.17 & 6.57 & 5.95 & b \\
171.4 & 136.8 & 177.6 & 178.8 & 179.6
 & 173.2 & 5.99 & 5.17 & 6.60 & 4.29 & b \\
171.4 & 136.8 & 177.6 & 178.8 & 179.7
 & 173.2 & 5.99 & 5.17 & 6.60 & 5.97 & b \\
5.3 & 172.9 & 2.3 & 174.8 & 0.5
 & 143.1 & 7.64 & 5.57 & 6.93 & 4.65 & b \\
5.3 & 172.9 & 2.3 & 174.8 & 0.3
 & 143.1 & 7.64 & 5.57 & 6.93 & 6.45 & b \\
5.5 & 172.9 & 2.2 & 174.8 & 0.4
 & 143.1 & 7.64 & 5.57 & 6.95 & 4.65 & b \\
5.5 & 172.9 & 2.2 & 174.8 & 0.3
 & 143.1 & 7.64 & 5.57 & 6.95 & 6.45 & b \\
174.5 & 172.9 & 178.1 & 174.8 & 179.6
 & 143.1 & 7.64 & 5.57 & 8.26 & 4.65 & b \\
174.5 & 172.9 & 178.1 & 174.8 & 179.7
 & 143.1 & 7.64 & 5.57 & 8.26 & 6.45 & b \\
174.7 & 172.9 & 178.1 & 174.8 & 179.5
 & 143.1 & 7.64 & 5.57 & 8.28 & 4.65 & b \\
174.7 & 172.9 & 178.1 & 174.8 & 179.7
 & 143.1 & 7.64 & 5.57 & 8.28 & 6.45 & b \\
\hline
\end{tabular}
\end{center} }
\caption{Output values of the strong and weak phases, as well as the
amplitudes, for values of the input parameters given in the text. All phase
angles are given in degrees. In the `Notes' column, `a', `b', `c' and
`a$^\prime$' indicate respectively the correct solution, a solution
inconsistent with other experimental constraints, a potential ambiguity,
and a solution with the correct values for the CP angles, but different
values for some of the other input parameters.}
\label{Sec6-Tab2}
\end{table}

The solutions can be obtained as follows. First, from
Eqs.~(\ref{DsA1}-\ref{DsB2}), we have
\bea
\cos\alpha &=& \frac{{\cal T}^2 + {\cal P}^2 - {A}}
 {2 {\cal T P}\cos\delta } 
 = \frac{\tilde{\cal T}^2 + {\tilde{\cal P}}^2 - \tilde{A}}
 {2 \tilde{\cal T} \tilde{\cal P}\cos\delta_s }~, 
\label{DsCA}\\
\sin\alpha &=& - \frac{B}{2 {\cal T P } \sin\delta }
 = - \frac{\tilde{B}}{2 \tilde{\cal T}\tilde{\cal P}\sin\delta_s }~, 
\label{DsSA}
\eea
and from Eqs.~(\ref{DsD1}-\ref{DsE2}), we have
\bea
\cos\gamma &=& \frac{r_u^2 {\cal T}^2 + F - D}
 {2 r_u {\cal T } \sqrt{F} \cos\delta^\prime }
 = \frac{r_u^2 \tilde{\cal T}^2 + \tilde{F} - \tilde{D}}
 {2 r_u \tilde{\cal T } \sqrt{\tilde{F}} \cos\delta_s^\prime },
\label{DsCG}\\
\sin\gamma &=& \frac{E}{2 r_u {\cal T} \sqrt{F}\sin\delta^\prime } 
 = \frac{\tilde{E}}{2 r_u \tilde{\cal T} 
 \sqrt{\tilde{F}}\sin\delta_s^\prime }. 
\label{DsSG}
\eea
The angles $\alpha$ and $\gamma$ can be eliminated from the above equations
to obtain the quadratic equations for $\cos\delta$, $\cos\delta^\prime$,
$\cos\delta_s$ and $\cos\delta^\prime_s$:
\bea
- 4 {\cal T}^2 {\cal P}^2 \cos^4\delta +
 \left\{ 4 {\cal T}^2 {\cal P}^2 - B^2 + 
 ( {\cal T}^2 + {\cal P}^2 - A )^2 \right\} \cos^2\delta 
~~~~~ & ~ & \nn \\
- ( {\cal T}^2 + {\cal P}^2 - A )^2 & = & 0 ~, 
\label{DsCD}\\
- 4 r_u^2 {\cal T}^2 F \cos^4\delta^\prime +
 \left\{ 4 r_u^2 {\cal T}^2 F - E^2 + 
 ( r_u^2 {\cal T}^2 + F - D )^2 \right\} \cos^2\delta^\prime 
~~~~~ & ~ & \nn \\
- ( r_u^2 {\cal T}^2 + F - D )^2 & = & 0 ~,
\label{DsCDp}\\
- 4 \tilde{\cal T}^2 \tilde{\cal P}^2 \cos^4\delta_s +
 \left\{ 4 \tilde{\cal T}^2 \tilde{\cal P}^2 - \tilde{B}^2 + 
 ( \tilde{\cal T}^2 + 
 \tilde{\cal P}^2 - \tilde{A} )^2 \right\} \cos^2\delta_s 
~~~~~ & ~ & \nn \\
- ( \tilde{\cal T}^2 + \tilde{\cal P}^2 - \tilde{A} )^2 & = & 0 ~,
\label{DsCDs}\\
- 4 r_u^2 \tilde{\cal T}^2 \tilde{F} \cos^4\delta_s^\prime +
 \left\{ 4 r_u^2 \tilde{\cal T}^2 \tilde{F} - \tilde{E}^2 + 
 ( r_u^2 \tilde{\cal T}^2 + 
 \tilde{F} - \tilde{D} )^2 \right\} \cos^2\delta_s^\prime 
~~~~~ & ~ & \nn \\
- ( r_u^2 \tilde{\cal T}^2 + \tilde{F} - \tilde{D} )^2 & = & 0 ~.
\label{DsCDsp}
\eea
Eliminating $\alpha$ and $\gamma$ in Eqs.~(\ref{DsC1}) and (\ref{DsC2}),
we also obtain the following equations:
\bea
4 {\cal P}^4 \left( B^2 + C^2 \right) \cos^4\delta -
 \left\{ 4 {\cal P}^2 B^2 ( {\cal T}^2 + {\cal P}^2 - A ) +
 4 {\cal P}^4 C^2 \right\} \cos^2\delta 
~~~~~ & ~ & \nn \\
+ ( {\cal T}^2 + {\cal P}^2 - A )^2 B^2 & = & 0 ~,
\label{DsC1CD}\\
4 \tilde{F}^2 \left( \tilde{E}^2 + \tilde{C}^2 \right) 
 \cos^4\delta_s^\prime -
 \left\{ 4 \tilde{F} \tilde{E}^2 
 ( r_u^2 \tilde{\cal T}^2 + \tilde{F} - \tilde{D} ) +
 4 \tilde{F}^2 \tilde{C}^2 \right\} 
 \cos^2\delta_s^\prime 
~~~~~ & ~ & \nn \\
+ ( r_u^2 \tilde{\cal T}^2 + \tilde{F} - \tilde{D} )^2 
 \tilde{E}^2 & = & 0 ~.
\label{DsC2CDsp}
\eea
We can therefore determine $\tilde{\cal T}$ and $\cos\delta^\prime_s$ from
Eqs.~(\ref{DsCDsp}) and (\ref{DsC2CDsp}). Then the angle $\gamma$ can be
obtained from Eq.~(\ref{DsSG}), and ${\cal T}$ and $\cos\delta^\prime$ from
Eqs.~(\ref{DsCG}) and (\ref{DsCDp}). For each ${\cal T}$, we can
determine ${\cal P}$ and $\cos\delta$ from Eqs.~(\ref{DsCD}) and
(\ref{DsC1CD}). This allows us to get the angle $\alpha$ from
Eq.~(\ref{DsSA}), and to determine $\tilde{\cal P}$ and $\cos\delta_s$ from
Eqs.~(\ref{DsCA}) and (\ref{DsCDs}). Evidently there are numerous
solutions, some of which can be eliminated by now reconsidering
Eqs.~(\ref{DsC1}) and (\ref{DsC2}), {\it i.e.} the signs of $C$
and $\tilde{C}$. Nevertheless, we are left with many possible solutions.

In Table \ref{Sec6-Tab2}, we give an example of these solutions. For the
amplitudes, we take ${\cal T} = 1, \tilde{\cal T} = 0.8$, ${\cal P}^\prime =
1, \tilde{\cal P}^\prime= 0.9$, ${\cal P} = \frac{|V_{us}|}{|V_{ud}|}
\frac{\sin\gamma }{\sin\alpha } {\cal P}^\prime$ and $\tilde{\cal P} = 1.1
{\cal P}$. The values assumed for the weak and strong angles
are $\alpha_{in} = 45.0^\circ$, $\gamma_{in} = 120.0^\circ $, $\delta_{in} =
84.3^\circ, \delta^\prime_{in} = 5.7^\circ, \delta_{s,in} =174.3^\circ$
and $\delta^\prime_{s,in} = 36.9^\circ$.

This table shows that there are 60 solutions for these equations (and there
are 60 others, not listed, in which $\pi$ is subtracted from all 6 output
angles). However, as before, not all solutions are allowed within the
context of the SM. In the table we have labeled the solutions as follows:
(a) correct solution, (b) one or more of the CP angles outside of the SM
domain, (c) potential ambiguity, and (a$^\prime$) correct solution for CP
angles but some of the other parameters are different from the inputs. As
is clear from the table, most of the spurious solutions disappear when one
imposes the SM constraints. However there are still some discrete
ambiguities which can not be eliminated.

There is one more point to make here. As explained previously, if the $u$-
and $c$-quark penguins are sizeable, then the CP angle $\alpha$ cannot be
obtained via this method. However, the new method for obtaining $\gamma$,
described in the previous section, is still viable, even when all four
strong phases are included. In this case the four measurements $\tilde{C}$,
$\tilde{D}$, $\tilde{E}$, and $\tilde{F}$, as obtained from the decays
$\bs(t) \to K^+ K^-$ and $\bs \to K^0 {\bar K^0}$, become:
\bea
\tilde{C} &=& \tilde{\cal T}^{\prime 2} \sin2\gamma - 2
\tilde{\cal T}^\prime \tilde{\cal P}_{\sss P}^\prime
\cos\delta_s^\prime \sin\gamma ~,
\label{gamC2} \\ 
\tilde{D} &=& \tilde{\cal T}^{\prime 2}+\tilde{\cal P}_{\sss P}^{\prime
2} - 2 \tilde{\cal T}^\prime \tilde{\cal P}_{\sss P}^\prime
\cos\delta_s^\prime \cos\gamma ~, 
\label{gamD2} \\ 
\tilde{E} &=& 2 \tilde{\cal T}^\prime \tilde{\cal P}_{\sss P}^\prime
\sin\delta_s^\prime \sin\gamma ~, 
\label{gamE2} \\ 
\tilde{F} &=& \tilde{P}_{\sss P}^{\prime 2} ~.
\label{gamF2}
\eea
Note that $\tilde{C}$-$\tilde{F}$ depend only on four parameters: 
$\tilde{\cal T}'$, $\tilde{\cal P}_{\sss P}^\prime$, $\delta'_s$,
and $\gamma$. Thus we have 4 equations in 4 unknowns, which is soluble. 
Thus, even in this worst-case scenario, where all corrections are
important, it is still possible to extract $\gamma$ from measurements of the
two processes $\bs(t) \to K^+ K^-$, and $\bs \to K^0 {\bar K^0}$. In this
case, there will be discrete ambiguities. 

\begin{table}
\begin{center}
\begin{tabular}{|c c |c|c|c|c|}
\hline 
$\gamma_{in}$ &$\delta_{s,in}^\prime$
 &$\gamma_{out}$ &$\delta_{s,out}^\prime$
 &$\tilde{\cal T}_{out}$ & Notes \\ \hline
60.0 & 36.9 & 60.0 & 36.9 & 0.80 & a \\
\ & \ & 47.5 & 8.3 & 3.93 & c \\
\ & \ & 141.9 & 171.7 & 4.69 & c \\
\ & \ & 171.7 & 143.1 & 4.78 & b \\
\hline
60.0 & 84.3 & 60.0 & 84.3 & 0.80 & a \\
\ & \ & 108.9 & 95.7 & 0.73 & c \\
\ & \ & 82.1 & 90.0 & 0.62 & c \\
\ & \ & 134.3 & 157.5 & 4.46 & c \\
\ & \ & 45.4 & 12.5 & 4.78 & c \\
\hline
60.0 & 174.3 & 60.0 & 174.3 & 0.80 & a \\
\ & \ & 124.9 & 178.9 & 4.40 & c \\
\ & \ & 42.8 & 1.1 & 5.31 & c \\
\ & \ & 5.84 & 5.7 & 6.82 & c \\
\hline
90.0 & 36.9 & 90.0 & 36.9 & 0.80 & a \\
\ & \ & 51.7 & 8.6 & 4.98 & c \\
\ & \ & 139.9 & 171.4 & 4.98 & c \\
\ & \ & 171.1 & 143.1 & 5.19 & b \\
\hline
90.0 & 84.3 & 90.0 & 84.3 & 0.80 & a \\
\ & \ & 46.7 & 14.4 & 4.41 & c \\
\ & \ & 134.8 & 165.6 & 4.52 & c \\
\ & \ & 128.5 & 95.7 & 1.02 & c \\
\hline
90.0 & 174.3 & 90.0 & 174.3 & 0.80 & a \\
\ & \ & 126.9 & 178.6 & 4.01 & c \\
\ & \ & 38.7 & 1.42 & 5.13 & c \\
\ & \ & 7.1 & 5.7 & 6.43 & b \\
\hline
120.0 & 36.9 & 120.0 & 36.9 & 0.80 & a \\
\ & \ & 53.6 & 6.8 & 4.39 & c \\
\ & \ & 136.8 & 173.2 & 5.17 & c \\
\ & \ & 172.9 & 143.1 & 5.57 & b \\
\hline
120.0 & 84.3 & 120.0 & 84.3 & 0.80 & a \\
\ & \ & 47.6 & 12.2 & 4.42 & c \\
\ & \ & 135.1 & 167.8 & 4.64 & c \\
\ & \ & 133.7 & 90.0 & 0.95 & c \\
\ & \ & 146.3 & 95.7 & 1.25 & c \\
\hline
120.0 & 174.3 & 120.0 & 174.3 & 0.80 & a \\
\ & \ & 131.9 & 178.6 & 3.77 & c \\
\ & \ & 35.8 & 1.4 & 4.80 & c \\
\ & \ & 6.6 & 5.7 & 6.02 & b \\
\hline 
\end{tabular}
\end{center}
\caption{Output values of $\gamma$, $\delta'_s$ and $\tilde{\cal T}$, for
given values of the input phases. All phase angles are given in degrees. In
the `Notes' column, `a', `b' and `c' indicate respectively the correct
solution, a solution inconsistent with other experimental constraints, and
a potential ambiguity.}
\label{Sec6-Tab3}
\end{table}

The solutions are obtained as in the previous cases. First, from
Eqs.~(\ref{gamD2}) and (\ref{gamE2}), we have 
\bea
\cos\gamma &=&\frac{\tilde{\cal T}^{\prime 2} + \tilde{F} - \tilde{D}}
 {2 \tilde{\cal T }^\prime \sqrt{\tilde{F}} \cos\delta_s^\prime }~,
\label{gamCG}\\
\sin\gamma &=& \frac{\tilde{E}}{2 \tilde{\cal T}^\prime 
 \sqrt{\tilde{F}}\sin\delta_s^\prime }~. 
\label{gamSG}
\eea
Eliminating $\gamma$ from Eqs.~(\ref{gamCG}) and (\ref{gamSG}) and in
Eq.~(\ref{gamC2}), we obtain the following quadratic equations
for $\cos\delta^\prime_s$: 
\bea
- 4 \tilde{\cal T}^{\prime 2} \tilde{F} \cos^4\delta_s^\prime +
 \left\{ 4 \tilde{\cal T}^{\prime 2} \tilde{F} - \tilde{E}^2 + 
 ( \tilde{\cal T}^{\prime 2} + 
 \tilde{F} - \tilde{D} )^2 \right\} \cos^2\delta_s^\prime 
~~~~~ & ~ & \nn \\
- ( \tilde{\cal T}^{\prime 2} + \tilde{F} - \tilde{D} )^2 & = & 0 ~,
\label{gamCDsp}\\
4 \tilde{F}^2 \left( \tilde{E}^2 + 
 \tilde{C}^2 \right) \cos^4\delta_s^\prime -
 \left\{ 4 \tilde{F} \tilde{E}^2 
 ( \tilde{\cal T}^{\prime 2} + \tilde{F} - \tilde{D} ) +
 4 \tilde{F}^2 \tilde{C}^2 \right\} \cos^2\delta_s^\prime 
~~~~~ & ~ & \nn \\
+ ( \tilde{\cal T}^{\prime 2} + \tilde{F} - \tilde{D} )^2 
 \tilde{E}^2 & = & 0 ~.
\label{gamC2CDsp}
\eea
Eqs.~(\ref{gamCG}-\ref{gamC2CDsp}) can now be solved straightforwardly to
give $\tilde{\cal T}^\prime$ and $\cos^2\delta^\prime_s$, and the CP
angle $\gamma$ can be obtained. As usual, there are multiple solutions,
some of which can be eliminated by now reconsidering Eq.~(\ref{gamC2}). 

In Table \ref{Sec6-Tab3}, we show examples of some of these solutions. For
the amplitudes, we take $\tilde{\cal T} = 0.8$ and $\tilde{\cal P}^\prime=
0.9$, and assume various values for the weak and strong phases, shown in
the table. This table shows that, for the values of the phases we have
chosen, there are always at least four solutions for these equations, some
of which are inconsistent with present experimental constraints ($30^\circ
\lsim \gamma \lsim 150^\circ$). (And there are other solutions, not listed,
in which $\{\gamma_{out}, \delta'_{s,out}\} \to \{\gamma_{out}-\pi,
\delta'_{s,out}\pi\}$.) In the table we have labeled the solutions as
follows: (a) correct solution, (b) the CP angles outside of the SM domain,
and (c) potential ambiguity. As is clear from the table, there are some
discrete ambiguities which can not be eliminated.

This method can also be applied to the $\Delta S = 0$ sector \cite{BF2}. The
analogue of $\bs(t) \to K^+ K^-$ is $\bd(t) \to \pi^+\pi^-$, so that this
technique might be a way of eliminating the troublesome penguin
contribution. However, things do not work quite as well for $\Delta S = 0$
decays. The main problem is that the only pure penguin decays are $B^+ \to
K^+ {\overline K^0}$ or $\bd \to K^0{\overline K^0}$, which at the quark
level are ${\bar b} \to {\bar d} s {\bar s}$. Within flavor $SU(3)$
symmetry, this is the same amplitude as ${\bar b} \to {\bar d} u {\bar u}$,
which contributes to $\bd \to \pi^+\pi^-$. However, $SU(3)$-breaking effects
are likely to ruin this equality, and it is very difficult to get an
accurate estimate of such effects. The analysis is also more complicated
in the $\Delta S = 0$ sector. Since the $u$- and $c$-quark contributions to
$b\to d$ penguins may be significant, this would lead to direct CP
violation in pure penguin decays in the $\Delta S = 0$ sector
\cite{Fleischer}. It is therefore necessary to perform a time-dependent
measurement of both $\bd(t) \to \pi^+\pi^-$ and $\bd(t) \to K^0{\overline
K^0}$ to disentangle all the parameters.


\section{Vanishing Strong Phases}

In the original DGR method, if $\delta=0$, the method breaks down, and
additional information is required to extract the CP angles. In this
section we examine what happens to the extended DGR method with $\bs$
decays if all strong phases vanish.

\begin{table}
\begin{center}
\begin{tabular}{|c c|r|r|r|r|l|c|}
\hline 
$\alpha_{in}$ &$\gamma_{in}$ & 
$\alpha_{out}$ &$\gamma_{out}$ &${\cal T}_{out}$ & 
$\tilde{\cal T }_{out}$ &${\cal P}_{out}$ & Notes \\ \hline
45.0 & 120.0 & 25.2 & 120.0 & 1.00 & 0.80 & 0.20 & c \\ 
\ & \ & 45.0 & 120.0 & 1.00 & 0.80 & 0.28 & a \\ 
\ & \ & 4.9 & 55.1 & 4.99 & 4.40 & 4.27 & b \\ 
\ & \ & 8.1 & 55.1 & 4.99 & 4.40 & 4.51 & b \\ 
\hline 
70.0 & 90.0 & 5.9 & 90.0 & 1.00 & 0.80 & 0.06 & b \\ 
\ & \ & 70.0 & 120.0 & 1.00 & 0.80 & 0.24 & a \\ 
\ & \ & 1.2 & 53.1 & 4.50 & 4.01 & 3.56 & b \\ 
\ & \ & 12.1 & 53.1 & 4.50 & 4.01 & 4.30 & b \\ 
\hline 
100.0 & 60.0 & 157.9 & 60.0 & 1.00 & 0.80 & 0.06 & b \\ 
\ & \ & 100 & 60.0 & 1.00 & 0.80 & 0.20 & a \\ 
\ & \ & 80.0 & 60.0 & 1.00 & 0.80 & 0.55 & c \\ 
\ & \ & 22.1 & 60.0 & 1.00 & 0.80 & 1.91 & b \\ 
\ & \ & 22.6 & 60.0 & 2.56 & 0.80 & 2.74 & b \\ 
\ & \ & 8.4 & 60.0 & 2.56 & 0.80 & 3.52 & b \\ 
\hline 
\end{tabular}
\end{center}
\caption{Output values of the weak phases and amplitudes for given values
of the input weak phases. All phase angles are given in degrees. In the
`Notes' column, `a', `b' and `c' indicate respectively the correct
solution, a solution inconsistent with other experimental constraints, and
a potential ambiguity.}
\label{Sec7-Tab4}
\end{table}

If $\delta=0$, Eqs.~(\ref{A1}-\ref{F1}) and (\ref{A2}-\ref{F2}) reduce to 8
equations in 8 unknowns: 
\bea
A &=& {\cal T}^2 + {\cal P}^2 - 2 {\cal T}{\cal P} \cos\alpha ~, 
\label{d=0,A1} \\
\tilde{A} &=& \tilde{\cal T}^2 + \tilde{\cal P}^2 
	- 2 \tilde{\cal T} \tilde{\cal P} \cos\alpha ~, 
\label{d=0,A2} \\
C &=& - {\cal T}^2 \sin2\alpha + 2 {\cal T P} \sin\alpha ~, 
\label{d=0,C1} \\
\tilde{C} &=& r_u^2 \tilde{\cal T}^2 \sin2\gamma 
 - 2 r_u \tilde{\cal T }\tilde{\cal P}^\prime \sin\gamma ~, 
\label{d=0,C2} \\ 
D &=& r_u^2 {\cal T}^2 + {\cal P}^{\prime 2} 
	- 2 r_u {\cal T}{\cal P}^\prime \cos\gamma ~, 
\label{d=0,D1} \\
\tilde{D} &=& r_u^2 \tilde{\cal T}^2 + \tilde{\cal P}^{\prime 2} 
	- 2 r_u \tilde{\cal T} \tilde{\cal P}^\prime \cos\gamma ~, 
\label{d=0,D2} \\ 
F &=& {\cal P}^{\prime 2} ~, 
\label{d=0,F1} \\
\tilde{F} &=& \tilde{\cal P}^{\prime 2} ~.
\label{d=0,F2}
\eea 
In this case, the equations can be solved for the 8 parameters. However,
numerical methods are required, and, as in the original DGR method
with $\delta\ne 0$, discrete ambiguities appear.

The solutions can be obtained as follows. First, from Eqs.~(\ref{d=0,A1})
and (\ref{d=0,A2}), we have
\bea
\cos\alpha = \frac{{\cal T}^2 + {\cal P}^2 - A }
 { 2 \cal T P } 
 = \frac{\tilde{\cal T}^2 + \tilde{\cal P}^2 - \tilde{A} }
 { 2 \tilde{\cal T} \tilde{\cal P} } ~,
\label{d=0,A1andA2}
\eea
and from Eqs.~(\ref{d=0,D1}-\ref{d=0,F2}),
\bea
\cos\gamma = \frac{ r_u^2 {\cal T}^2 + F - D }
 { 2 r_u {\cal T} \sqrt{F} } 
 = \frac{ r_u^2 \tilde{\cal T}^2 + \tilde{F} - \tilde{D} }
 { 2 r_u \tilde{\cal T} \sqrt{\tilde{F}} } ~. 
\label{d=0,D1andD2}
\eea
Eliminating $\alpha$ and $\gamma$ in Eqs.~(\ref{d=0,C1}) and (\ref{d=0,C2}),
we obtain the following equations: 
\bea
C^2 &=& \left( -2 {\cal T}^2 \frac{{\cal T}^2 + {\cal P}^2 - A }
 { 2 \cal T P } + 
 2 {\cal T P } \right)^2 
 \left( 1 - \frac{({\cal T}^2 + {\cal P}^2 - A )^2}
 { 4 {\cal T}^2 {\cal P}^2 } \right), 
\label{d=0,no sin a} \\
\tilde{C}^2 &=& \left( 2 r_u^2 \tilde{\cal T}^2 
 \frac{ r_u^2 \tilde{\cal T}^2 + \tilde{F} - \tilde{D} }
 { 2 r_u \tilde{\cal T }\sqrt{\tilde{F}} } - 
 2 r_u \tilde{\cal T } \sqrt{\tilde{F}} \right)^2 
\left( 1 - \frac{( r_u^2 \tilde{\cal T}^2 + \tilde{F} - \tilde{D})^2}
 { 4 r_u^2 \tilde{\cal T }^2 \tilde{F} }\right).
\label{d=0,no sin g}
\eea
We can therefore determine $\tilde{\cal T}$ from Eq.~(\ref{d=0,no sin g}).
This allows us to get ${\cal T}$ from Eq.~(\ref{d=0,D1andD2}), and
then ${\cal P}$ from Eq.~(\ref{d=0,no sin a}). The angles $\alpha$
and $\gamma$ can then be obtained from Eqs.~(\ref{d=0,A1andA2}) and
(\ref{d=0,D1andD2}). There are, of course, many solutions. Some of these
can be eliminated by reconsidering Eqs.~(\ref{d=0,C1}) and (\ref{d=0,C2}), 
{\it i.e.} the signs of $C$ and $\tilde C$, but multiple solutions still
remain.

To illustrate this, we take ${\cal T } = 1$, $\tilde{\cal T} = 0.8$, ${\cal
P}^\prime = 1$, $\tilde{\cal P}^\prime= 0.9$, ${\cal P} =
\frac{|V_{us}|}{|V_{ud}|} \frac{\sin\gamma }{\sin\alpha } {\cal P}^\prime$,
$\tilde{\cal P} = 1.1 {\cal P}$. We choose three representative sets of
values of the CP angles $\alpha$ and $\gamma$, and solve the equations as
described above. The solutions are shown in Table \ref{Sec7-Tab4}.

As is clear from the table, there are many ambiguities (and there are
additional solutions in which $\{\alpha_{out}, \gamma_{out}\}$ are replaced
by $\{\alpha_{out}-\pi, \gamma_{out}-\pi\}$). As before, not all solutions
are allowed within the context of the SM. We have labeled the solutions as
(a) correct solution, (b) one or more of the CP angles outside of the SM
domain, and (c) potential ambiguity. For the particular values of $\alpha$
and $\gamma$ that we have chosen, there are a few solutions consistent with
the SM. 

Finally, we note that, for the special case in which $\tilde{\cal T} =
{\cal T}$, $\tilde{\cal P} = {\cal P}$, and $\tilde{\cal P}^\prime = {\cal
P}'$, the system reduces to 5 equations in 5 unknowns, which can be solved
just as above. One can still extract the parameters up to discrete
ambiguities.


\section{Conclusions}

The method proposed by Dighe, Gronau and Rosner (DGR) for obtaining the CP
angles $\alpha$ and $\gamma$ involves the measurement of the decays
$\bd(t)\to\pi^+\pi^-$, $\bd \to \pi^- K^+$, $B^+ \to \pi^+ K^0$, and their
charge-conjugate processes, and assumes $SU(3)$ flavor symmetry and
first-order $SU(3)$ breaking. This method has a number of advantages: there
are no problems with electroweak penguins, all decays are accessible at
asymmetric $e^+e^-$ $B$ factories, and the decays involve only charged
$\pi$'s or $K$'s, which are easy to detect experimentally. Even so, there
are some problems as well. First, there are a large number of discrete
ambiguities in the extraction of the CP angles. Second, some theoretical
assumptions are required: the strong phases in $\Delta S=0$ transitions are
assumed to be equal to their counterparts in the $\Delta S=1$ transitions,
even in the presence of $SU(3)$ breaking, and the $b\to d$ penguin is
assumed to be dominated by an internal $t$ quark. If either of these
assumptions is relaxed, then there is not enough information to determine
the CP angles. Finally, if all strong phases vanish, the method again
breaks down.

In this paper, we have proposed an extension of this method which avoids
most of the problems with the DGR method. In addition to the $\bd$ and
$B^+$ decays used by DGR, it requires the measurement of their
$SU(3)$-counterpart $\bs$ decays: $\bs \to \pi^+ K^-$, $\bs(t) \to K^+ K^-$,
and $\bs \to K^0 {\bar K^0}$. This overconstrains the system, which
eliminates most discrete ambiguities in the extraction of $\alpha$
and $\gamma$. Furthermore, if DGR's assumptions are relaxed, there is still
enough information in most cases to obtain the CP angles.\\

We have found the following results:
\begin{enumerate}

\item If we make the same assumptions as DGR, we are able to extract
$\alpha$ and $\gamma$ up to a 2-fold ambiguity, corresponding to the
unitarity triangle pointing up or down. In the special case where the
magnitudes of the amplitudes are independent of the spectator quark, it is
still possible to extract the CP angles up to the same 2-fold ambiguity.

\item If we allow the strong phases in the $\Delta S=1$ transitions to be
different from those in the $\Delta S=0$ transitions, we can still
obtain $\alpha$ and $\gamma$ with a 2-fold ambiguity. If the magnitudes of
the amplitudes are independent of the spectator quark, then it is still
possible to extract the CP angles, but there are multiple discrete
ambiguities. Many of these can be eliminated by imposing the present
experimental constraints on the angles, but of course one might be
overlooking the presence of new physics by doing so.

\item If we consider nonzero$u$- and $c$-quark contributions to the $b\to d$
penguin, then, strictly speaking, the method partially breaks down. The
angle $\alpha$ cannot be extracted, but $\gamma$ can still be obtained up to
a 2-fold ambiguity. However, if the $u$- and $c$-quark penguins are much
smaller than the tree diagram (even if they are not negligible compared to
the $t$-quark penguin) then this situation reduces to the case of different
phases in the $\Delta S=0$ and $\Delta S=1$ transitions, described above. So
it may still be possible to obtain $\alpha$ in this case, but some
theoretical uncertainty may be introduced.

\item When one includes $\bs$ decays, there are two distinct ways in which
$SU(3)$ can be broken: (i) $\Delta S=0$ processes {\it vs.} $\Delta S=1$
processes, and (ii) $\bd$ decays {\it vs.}$\bs$ decays. Therefore, if one
includes all types of first-order $SU(3)$ breaking, four different strong
phases must be considered. In this case, we find that the CP
angles $\alpha$ and $\gamma$ can be extracted, but with multiple discrete
ambiguities.

\item If all strong phases vanish, then one can still obtain $\alpha$ and
$\gamma$, up to multiple discrete ambiguities.

\end{enumerate}

Finally, we have found a new method of measuring $\gamma$. By
measuring $\bs(t) \to K^+ K^-$ and $\bs \to K^0 {\bar K^0}$, it is possible
to obtain $\gamma$, up to discrete ambiguities, with no hadronic
uncertainties. Experimentally this will be difficult, as it requires
isolating the tree contribution to $\bs \to K^+ K^-$ by ``subtracting off"
the penguin contribution. However, this penguin contribution is much
larger than the tree, which means that one is essentially subtracting two
big numbers to get a small number. Still, $B$-physics experiments at hadron
colliders may have the precision to carry out such measurements.\\

\bigskip
\centerline{\bf Acknowledgements}
\bigskip

C.S.K. thanks D.L. for the hospitality of the Universit\'e de Montr\'eal,
where some of this work was done. The work of C.S.K. was supported in part
by the CTP of SNU, in part by the BSRI Program, BSRI-97-2425, in part by
the KOSEF-DFG, 96-0702-01-01-2 and in part by Yonsei University Research Fund
of 1997. The work of D.L. was financially supported
by NSERC of Canada and FCAR du Qu\'ebec. The work of T.Y. was supported in
part by the Japan Society for the Promotion of Science.

\end{document}